\documentclass[11pt,twoside]{article}
\usepackage{cozumel2005}
\usepackage{epsf}
\usepackage{psfig}
\usepackage{lscape}
\usepackage{wasysym}

\begin{document}

\title{Populations of AGB stars and LPVs in the Galaxy and Local Group} 

\author{M.A.T. Groenewegen }   

\affil{Instituut voor Sterrenkunde \\
Celestijnenlaan 200B \\
B-3001 Leuven, Belgium
}    

\begin{abstract} 
In this review I will discuss recent developments on the topic of
resolved Asymptotic Giant Branch (AGB) stars and Long Period Variables
(LPVs) in our Galaxy, the Magellanic Clouds, and the Local Group in
general. Since the main characteristics of AGB stars are (1) their
infrared, (2) pulsational and (3) chemical properties, it is fitting
that I will focus on recent results from the (a) 2MASS/DENIS and MSX
near- and mid-IR surveys, the (b) micro-lensing surveys and (c)
dedicated narrow-band imaging surveys.
The presentation is available at http://www.ster.kuleuven.be/$^{\sim}$groen

\end{abstract}


\section{Introduction}                      

All main-sequence stars born with masses below $\la$ 8 $M_{\odot}$
have or will go through the evolutionary phase called Asymptotic Giant
Branch (AGB). The lower limit in initial mass is set by the age of the
Galactic Disc, the upper limit is set by the mass where carbon can be
ignited in the stellar core. The AGB is the final phase where
intermediate-mass stars have nuclear burning in the form of alternate
Hydrogen and Helium shell burning, before they cross the
Herzsprung-Russell diagram to become Planetary Nebulae and then White
Dwarfs.

A summary of the interior structure and stellar evolution up to and
on the AGB can be found in the recent textbook ``Asymptotic Giant
Branch Stars'' (Habing \& Oloffson 2004).

AGB stars are luminous ([$\sim$0.1 -- a few] 10$^4$ $L_{\odot}$) and
cool, with effective temperatures in the range 3850 to $\sim$2500 K
(for M0 to M10 giants, e.g. Fluks et al. 1994). From this it follows
that AGB stars are big (up to a few hundred $R_{\odot}$), and
combining this with the classical pulsation equation\footnote{With the
pulsation constant $Q$ being of order 0.038.}  $P = Q \; R^{1.5} \; M^{-0.5}$
it follows that any radial pulsations that would occur would have
periods between tens and hundreds of days.

Equally important, and typical for the AGB, are the chemical
peculiarities that occur during this evolutionary phase (see Chapter 2
in the abovementioned book). Depending in a complex way on initial mass,
metallicity, mass-loss, mixing and burning in the envelope [hot
bottom-burning], an AGB star may go through several third dredge-up
events whereby mainly carbon, nitrogen, oxygen and $s$-process
elements are mixed ultimately into the stellar photosphere. Depending
on the C/O-ratio different molecules form in the cool atmospheres
(VO, TiO, C$_2$, CN) and a star can be classified as M-star (C/O $\la$
0.95), S-star (0.95 $\la$ C/O $<$ 1.0) or C-star (C/O $\ge$ 1.0). 
Intermediate classes, MS, SC, also exist. These type of AGB C-stars
are also identified with the N-type carbon stars, and they are
different from the R-type carbon stars which are likely not on the
AGB, and possibly related to a merger of stars on the RGB (see McClure 1997).

The low effective temperatures already make AGB stars redder than all
their Main-Sequence progenitors. In addition, the formation of
different molecules depending on chemical type makes that the infrared colours
of M- and C-stars are different, as will be discussed later.

Furthermore for spectral types later than $\sim$M4-M5 (e.g. Glass \&
Schultheis 2002) the region close to the star has the right
combination of temperature and density for dust grains to form. Dust
absorbs efficiently in the optical and radiates in the infra-red. This
implies that AGB stars surrounded by dust shells are even redder.

So far, the binary chanel has not been discussed. Main-sequence,
sub-giants and giants that have been polluted when a present-day WD
was on the AGB and polluted his companion with material enriched in
carbon and $s$-process elements. Examples of these classes are the
carbon-dwarfs (e.g. Steinhardt \& Sasselov 2005), the CH-stars
(e.g. Bartkevicius 1996, McClure \& Woodsworth 1990), the Barium stars
(e.g. Jorissen et al. 1998) and the ``extrinsic'' S-stars (e.g. Van
Eck \& Jorissen, 1999). Recently, it has become clear that about 20\%
of Carbon-enhanced Extreme Metal-Poor (CEMP) stars are binaries and
show s-process enhancements (Lucatello et al. 2005). These classes
will not be discussed further here.

Based on the main properties of AGB stars mentioned above, the
following recent results will be discussed: 2MASS/DENIS and MSX in
connection with the infrared colours, the result of the micro-lensing
surveys in connection with variability, and the results of dedicated
narrow-band surveys in connection with the chemical characteristics of
AGB stars.

\section{AGB stars in the 2MASS/DENIS  and MSX surveys}

\subsection{2 Micron All Sky Survey -- DEep Near-Infrared Survey}

In May 2003 the DENIS team released their second data release
containing $IJK$ photometry for 195 million point sources in the
southern sky down to $K$ = 14.0 (available at VIZIER at
http://vizier.u-strasbg.fr/viz-bin/VizieR?-source=B/denis)

In March 2003 the final {\it all-sky} 2MASS $JHK$-database was
released containing 470 million point sources down to a typical
limiting magnitude of $K$ = 14.3 (Cutri et al. 2003).  This implies
that a typical carbon star {\it even with no mass loss} and with a
luminosity of 7000 L$_\odot$ and effective temperature of 2650 K can
be viewed to 150 kpc, and mass-losing AGB stars to even larger
distances.

This suggests that AGB stars in many Local Group galaxies have been
detected by 2MASS. Groenewegen (2005) presents a list of almost 100
candidates in 16 Local Group (LG) galaxies (excluding the Magellanic
Clouds, M31, M33), selected on $(J-K)$-colour and $K$-magnitude. We
hope to confirm the 34 candidates in Fornax dSph by NIR spectroscopy
in the fall of 2005 using the VLT.

In the near future much deeper NIR survey data of LG galaxies is
within reach of wide-field imaging facilities as UKIDSS
(http://www.ukidss.org) and VISTA (http://www.vista.ac.uk).  \\

\begin{figure}[!t]
\plottwo{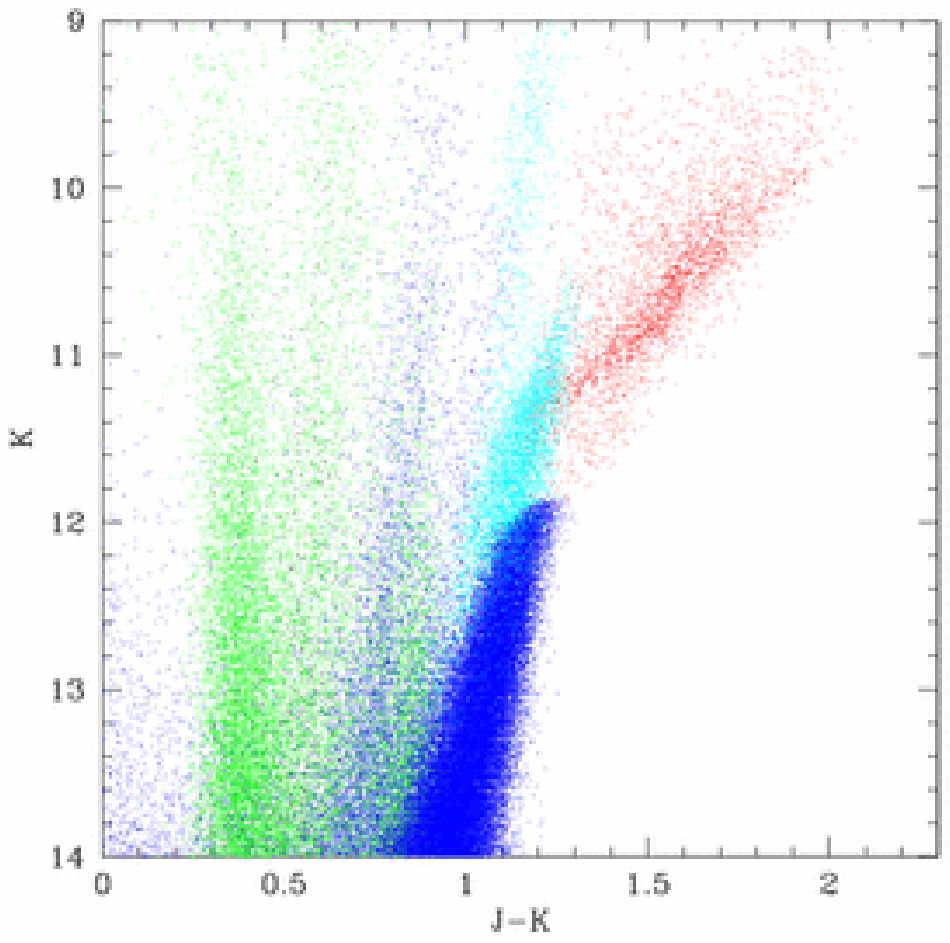}{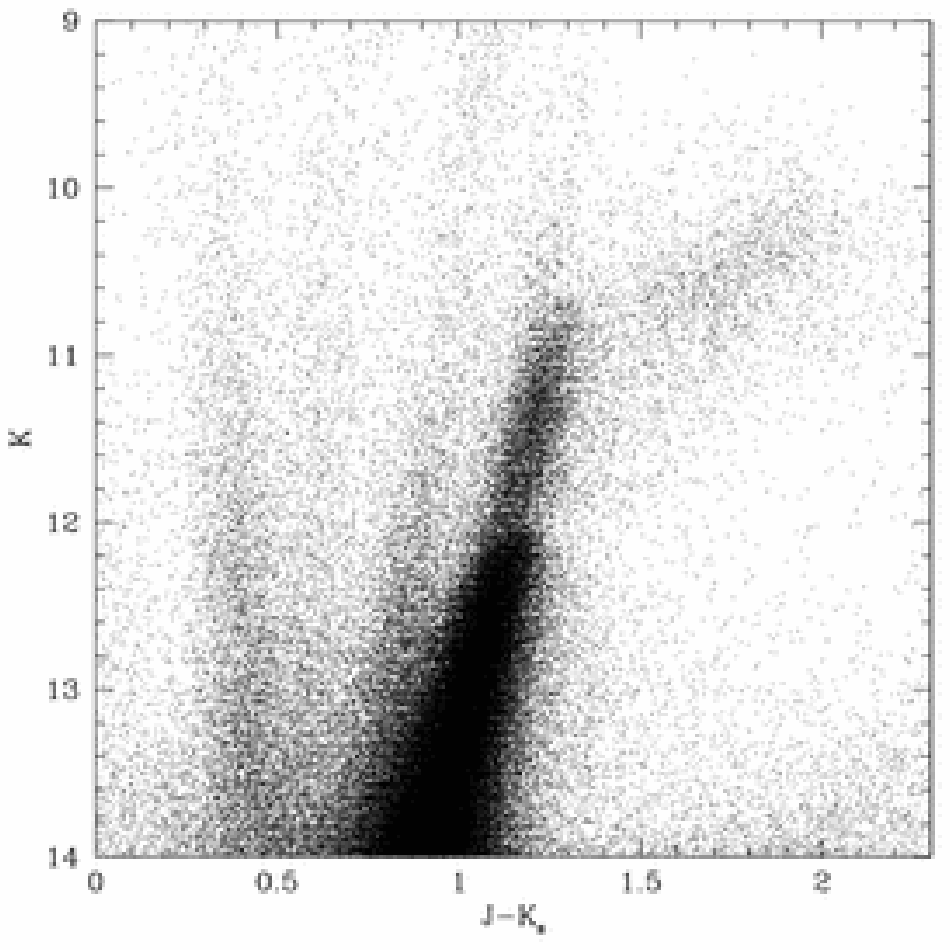}
\caption[]{
From Marigo et al. (2003; their figure~12). Left panel: Simulated CMD
based on TP-AGB tracks with variable molecular opacities (i.e. Marigo
2002), assuming an half-and-half mixture of fundamental-mode and
first-overtone pulsators for the stars which evolve into the C-star
phase. C-stars are marked in red and O-rich low-mass TP-AGB stars in cyan.
Green is foreground, and dark blue are RGB stars.
Right panel: The same diagram for the selected 2MASS data. Note the
good correspondence between the different featured in the CMD.
}
\label{Fig-D}
\end{figure}

\noindent
Up to now, the 2MASS and DENIS data have extensively been used to
study AGB stars in the Magellanic Clouds:

\begin{itemize}

\item Weinberg \& Nikolaev (2001) who select stars with $9.5 \la K \la
  11.5$ and $1.4 \la (J-K) <2$ (assumed to be C-rich LPVs) to study
  the 3D structure.

\item Cioni \& Habing (2003) present results on the C/M ratio across the
face of the SMC and LMC (results on NGC 6822 are presented in Cioni \&
Habing, 2005) using DENIS data. Carbon and O-rich AGB stars are
selected from the $(I-J)-(J-K)$ colour-colour (CC)-diagram. The C/M
ratio is then determined in 100 $\times$ 100 cells of 0.04 square
degrees. In the SMC there is no clear trend with position while in the
LMC the ratio appears to decrease radially. Overall the distribution
is clumpy and corresponds to a spread in metallicity of 0.75 dex.

\item Marigo et al. (2003; see Figure~\ref{Fig-D}) fit the
colour-magnitude diagram (CMD) of 2MASS objects within 4 degrees of
the bar optical centre using a population synthesis code (Girardi et
al. 2005), with special emphasis on the tail of carbon stars. She
demonstrates the importance of including the change in molecular
opacities in evolutionary calculations as stars evolve from O-rich to
C-rich (Marigo 2002).

\end{itemize}

\noindent
Regarding our own Milky-Way it is interesting to mention the Faint
High-Latitude Carbon (FHLC) stars\footnote{The halo usually defined as
$\mid b \mid \ga 30$ degrees in this context.}. Originally one was
interested in identifying carbon stars at high galactic latitude as
tracers of the halo. However, quite a few turned out to have a
measurable parallax or proper motion, indicating that these were not
giants, but rather carbon dwarfs. Papers that use 2MASS data in
connection with FHLC stars are:

\begin{itemize}

\item
Liebert et al. (2000) looked for very red objects in the halo in the
2MASS second incremental data release and confirmed, taking optical
spectra, four new halo carbon stars with $(J-K)$ between 2.3 and 4.7.
These stars are similar to two examples found earlier using IRAS
(e.g. Groenewegen et al., 1997, and references therein).

\item
Mauron et al. (2004) did a similar study using the 2MASS but their initial
selection was based on $(H-K)>0.4$ and $(J-H)> 0.95$, hence
significantly bluer than the stars found by Liebert et al., but not as
blue as some carbon dwarfs. Their initial sample of about 1200 objects
was trimmed by eliminating known objects using SIMBAD and stars with
$(B-R)<1.5$ from the USNO.  They ended up with a list of $\sim$200
best candidates and optical spectra were taken of 97 and 30 were
confirmed as carbon stars.

\item
Downes et al. (2004) follow up on the work by Margon et al. (2002) by
selecting FHLC stars using 5-colour Sloan Digital Sky Survey (SDSS)
photometry, and then taking optical spectra. Correlation with the
2MASS is done to provide the IR magnitudes. Margon et al. find 39, and
Downes et al. 251 FHLC stars. At least half of these are dwarf carbon
stars. The density of FHLC stars is about 0.06 per square degree down
to $r^* = 21$. \\

\end{itemize}

\subsection{Midcourse Space Experiment}

The {\it Midcourse Space Experiment} (MSX; see Price et al. 2001)
satellite mission included the {\sc spirit iii} instrument that
obtained mid-IR data in 6 bands designated A,B1,B2,C,D,E with central
wavelengths between 8 and 21 $\mu$m.  Band A was the most sensitive one
with a limiting flux-density of order 0.1 Jy, and most of the 320~000
objects are only detected in this band.

MSX observed the Galactic plane ($\mid b \mid <5\deg$) and the LMC.
Lumsden et al. (2002) and Ortiz et al. (2005) investigate the colours
of AGB stars. Both papers reach similar conclusions in that the
combination of MSX magnitudes with near-infrared magnitudes ($J$ or $K$) 
is a powerful diagnostic in discriminating between different types
of reddened sources.

The MSX results on the LMC are discussed in Egan et al. (2001).  1806
sources were detected in the $A$ band of which 1664 had positional
matches with 2MASS. This allowed them to discriminate different
classes of objects in a ($J-K$)-($K-A$) CC-diagram.

\begin{table}[!t]
\caption{Red variables from micro-lensing surveys}
\small
\setlength{\tabcolsep}{1.4mm}

\begin{tabular}{llll} \hline
Reference  & Area & Number & survey data \\
\hline
Wood et al. 1999 \&         &       &                     &  \\
$\;\;\;$Wood 2000           & 0.25 $\Box^2$ LMC-bar & 1430 RV$^{(1)}$  & MACHO + IR \\ 
Cioni et al. 2001           & 0.5 $\Box^2$  LMC-OC  & 240 M+SR & EROS + D$^{(2)}$ \\
Noda et al. 2002            & 14 $\Box^2$ LMC       & 146 LPV & MOA + D \\ 
Lebzelter et al. 2002       & 0.25 $\Box^2$ LMC-bar & 470 RV & AGAPEROS + D \\ 
Cioni et al. 2003 & 0.25 $\Box^2$ SMC & 458 RV$^{(4)}$ & MACHO+D/2M$^{(3)}$ \\
Ita et al. 2004a,b          & 3 $\Box^2$ LMC-centre & $\sim$9000 RV &  OGLE + SIRIUS \\
                            & 1 $\Box^2$ SMC-centre & $\sim$3000 RV &  OGLE + SIRIUS \\
Kiss \& Bedding 2003        & 4.5 $\Box^2$ LMC-centre & $\sim$23000 RV & OGLE + 2M  \\ 
                            &                    &                     & with $J-K >0.9$ \\
Kiss \& Bedding 2004        & 2.5 $\Box^2$ SMC-centre & $\sim$3200 RV & OGLE + 2M  \\ 
                            &                   &                     & with $J-K >0.9$ \\
Groenewegen 2004            & SMC+LMC & 2277 SC$^{(5)}$ & OGLE + D/2M \\
Soszynski  et al. 2004      & SMC+LMC & 18~000 SARV$^{(6)}$ & OGLE \\
Noda et al. 2004            & LMC     & 4000 RV         & MOA + D \\ 
Fraser et al. 2005          & LMC     & 22~0000 RV      & MACHO + 2M \\
Raimondo et al. 2005        & SMC     & 1080 C-stars    & MACHO + 2M/D \\
 & & & \\
Alard et al. 2001        & GB (NGC6522+Sgr I) & 332 & MACHO+ IG$^{(7)}$ \\

Glass \& Schultheis 2002 & GB (NGC6522) & 174 M-giants & MACHO+D+IG \\

Glass \& Schultheis 2003 & GB (NGC6522) & 1085 RV & MACHO + D \\

Wray et al. 2004         & GB  & 13~000 SARV & 33 OGLE-fields \\

GB05                     & GB & 2691 Miras & OGLE + D/2M \\

\hline
\end{tabular}

(1) RV = red variables.
(2) D = DENIS $IJK$ survey.
(3) 2M = 2MASS $JHK$ survey.
(4) A pre-selected sample of stars detected in an ISOCAM survey (Loup et al., in prep.).
(5) SC = Spectroscopically Confirmed  stars.
(6) SARV = small-amplitude red variables.
(7) IG= ISOGAL mid-IR survey.

\end{table}

\section{Long Period Variables in the micro-lensing surveys}

Before focusing on the results of the micro-lensing surveys {\it per
se} on LPV research, it is worthwhile to mention a few variability
surveys that also are in the process of providing results on variable
stars.  Contrary to the micro-lensing surveys the surveys mentioned
below will be most useful for studies of variable stars in the solar
neighbourhood and a welcome complement to the classical General
Catalog of Variable Stars (GCVS, Kholopov et al., 1998) which lists
7200 Mira variables. These new surveys are:

\begin{itemize}

\item ASAS (All-Sky Automated Survey), see Pojmanski (2002).

This survey monitored about 15 million stars in the magnitude range
$V = 5 - 14$, south of declination +28 degrees, for 3 years every 1-3 days.

Of the 38~000 variables in the ASAS-3 release 2551 are classified as
Mira variables on the web-site.

\item NSVS (Northern Sky Variability Survey), see Wo\'zniak et al. (2004a).

This survey monitored about 14 million objects in the northern sky
(and some fields with less quality down to $-38$ in declination), in
the optical magnitude range 8 to 15.5, over a 1 year baseline with 100-500
measurements per object.

Wo\'zniak et al. (2004b) use this dataset to create a catalog of 8600
variable AGB stars.

\item TAROT (Klotz, this proceedings)

\end{itemize}

\begin{figure}[!t]
\begin{minipage}{0.49\textwidth}
\resizebox{\hsize}{!}{\includegraphics{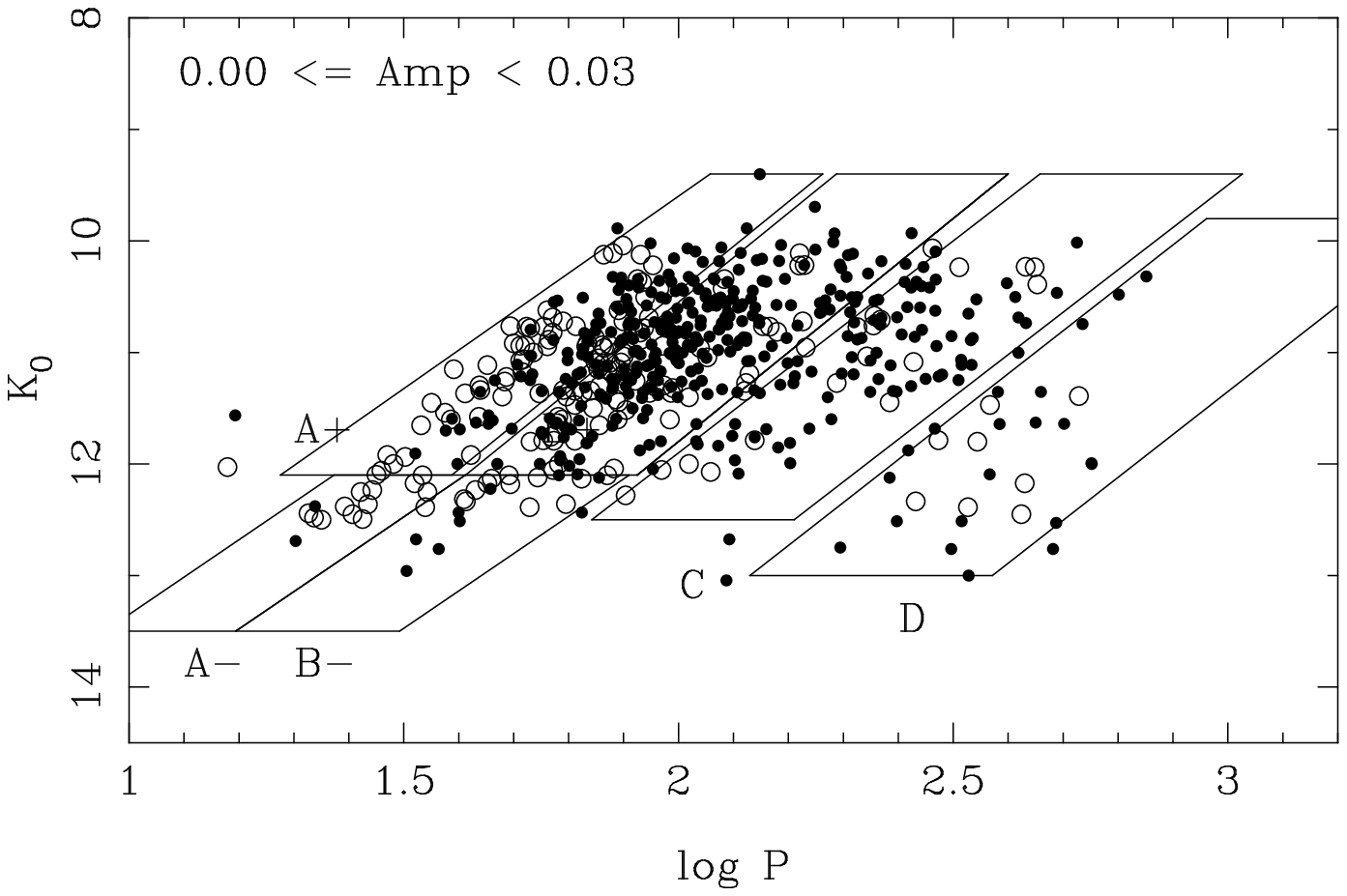}}
\end{minipage}
\hfill
\begin{minipage}{0.49\textwidth}
\resizebox{\hsize}{!}{\includegraphics{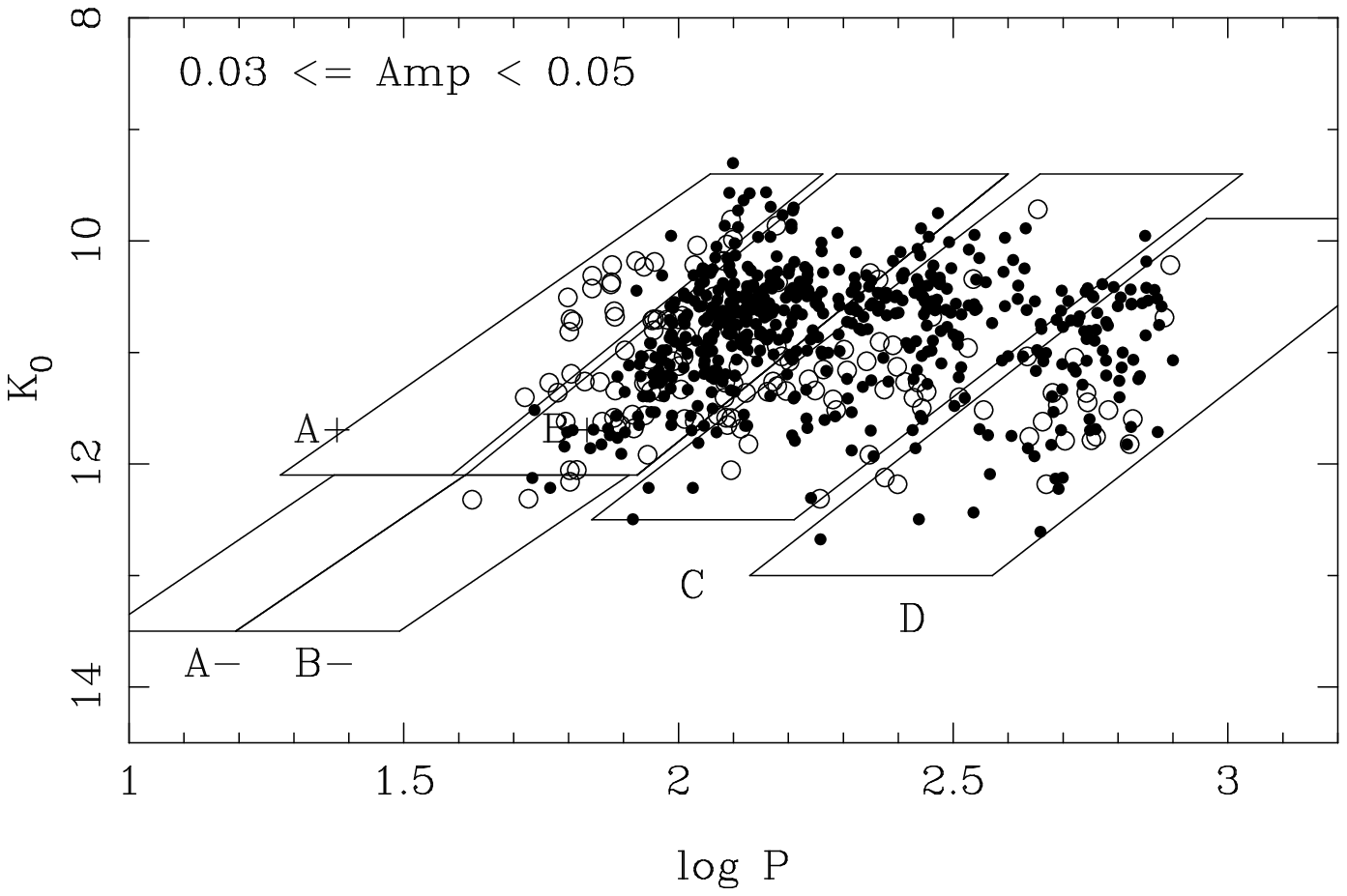}}
\end{minipage}

\begin{minipage}{0.49\textwidth}
\resizebox{\hsize}{!}{\includegraphics{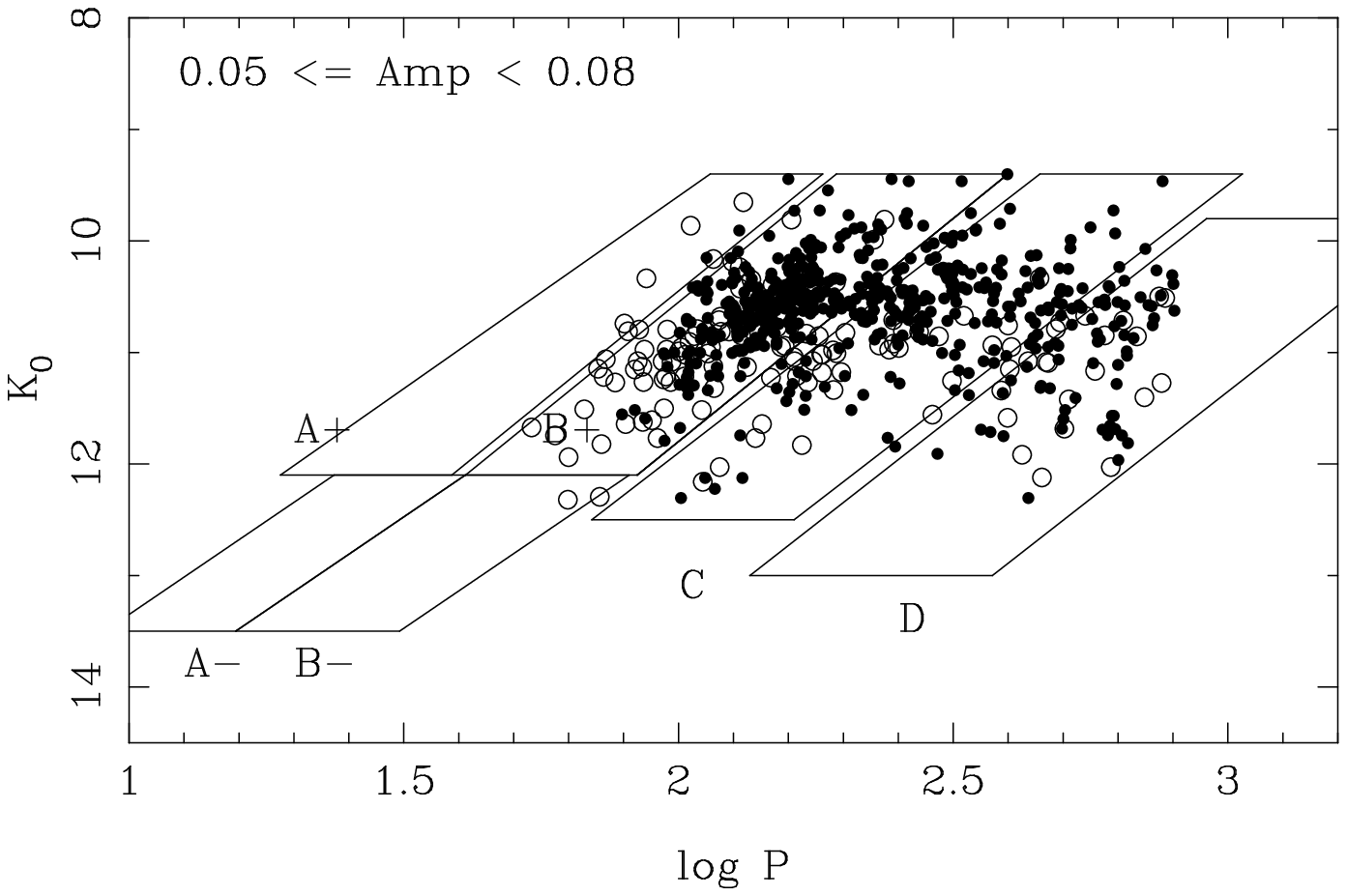}}
\end{minipage}
\hfill
\begin{minipage}{0.49\textwidth}
\resizebox{\hsize}{!}{\includegraphics{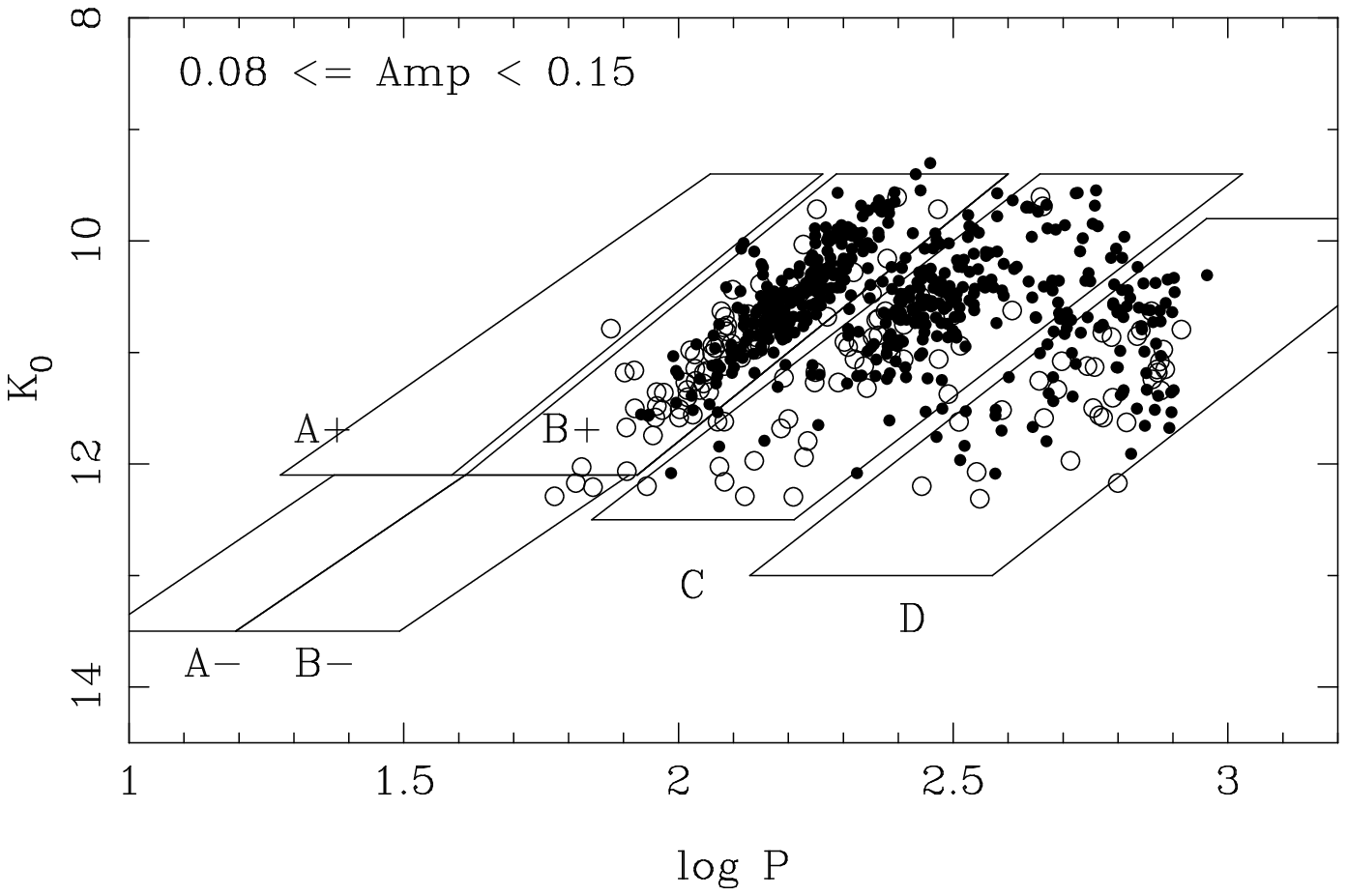}}
\end{minipage}

\begin{minipage}{0.49\textwidth}
\resizebox{\hsize}{!}{\includegraphics{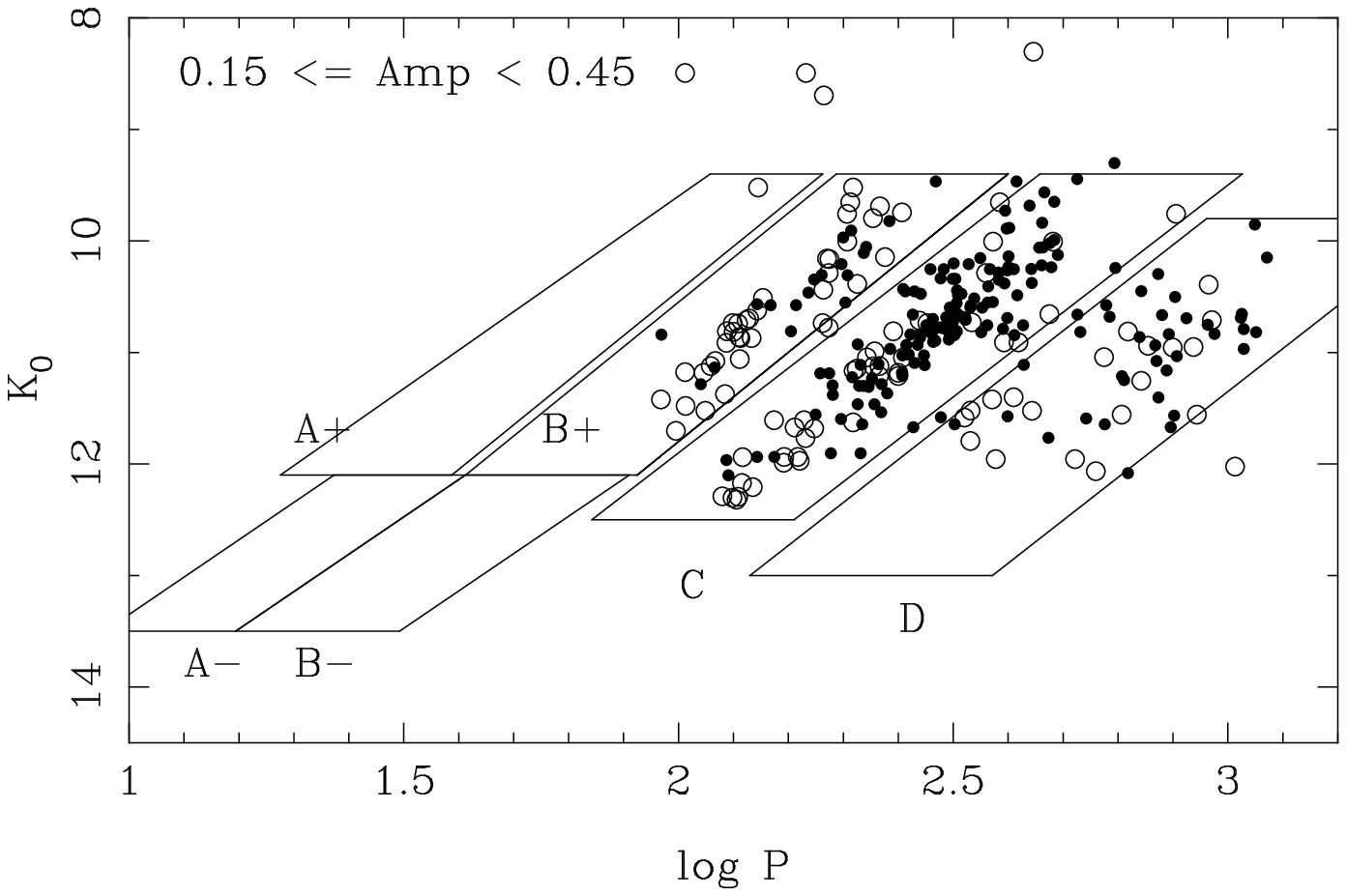}}
\end{minipage}
\hfill
\begin{minipage}{0.49\textwidth}
\resizebox{\hsize}{!}{\includegraphics{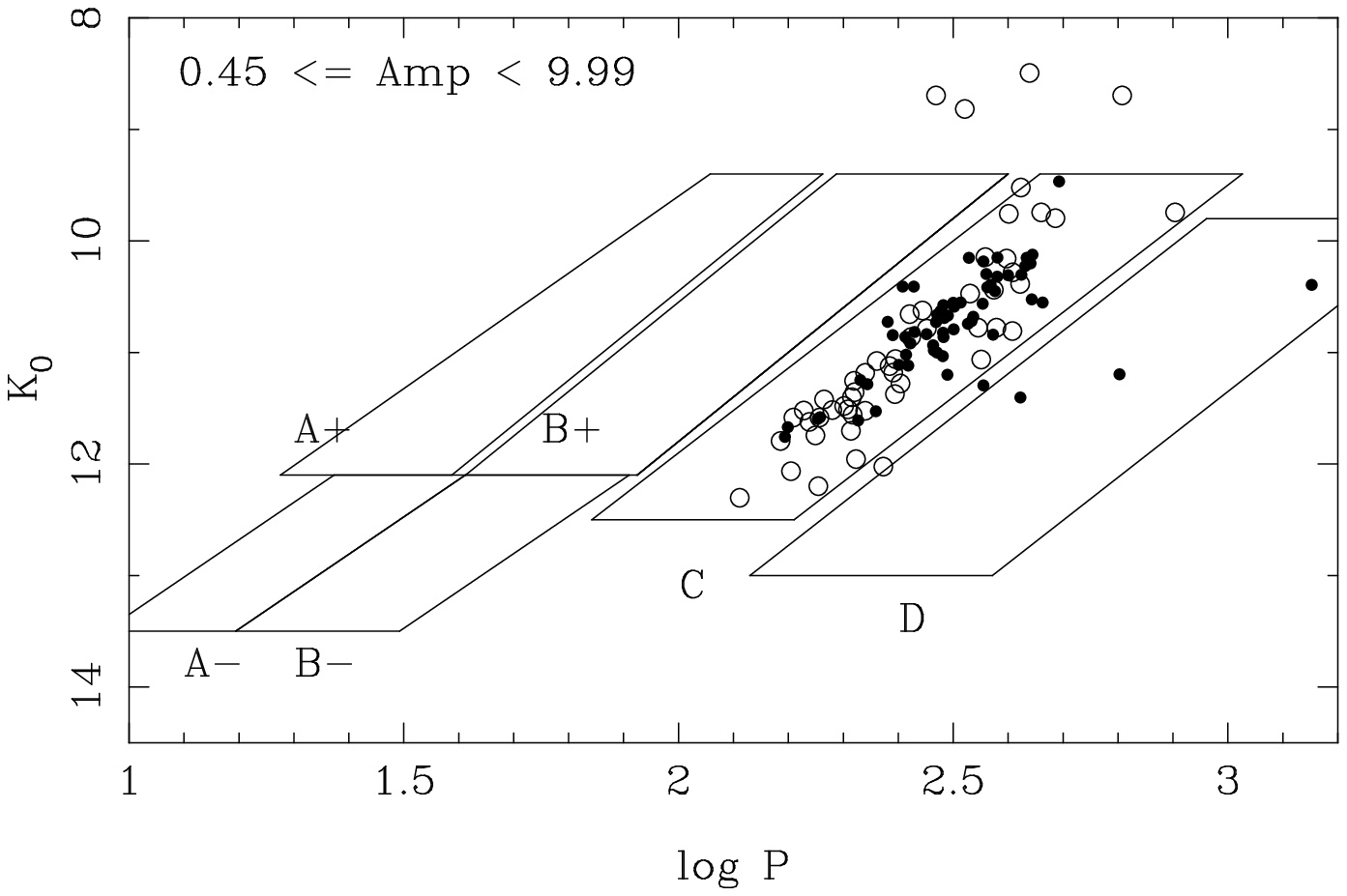}}
\end{minipage}
\caption{
$K$-band $PL$-relation for the LMC. Panels indicate selection on
$I$-band amplitude as indicated in the upper left corners.  Carbon
stars are indicated by filled circles, M- and S-stars by open
circles. Boxes related to the ``ABCD'' sequences are indicated. From
Groenewegen (2004).
}
\label{fig-PL-MC}
\end{figure}

\subsection{The Magellanic Clouds}

Lets us now turn to the micro-lensing surveys. After the initial
results presented by Wood et al. (1999) and Wood (2000) on a single
field of MACHO data in the LMC, there has been a flood of papers on
this topic. Table~1 summarises the current state of affaires.  The main
differences are in the location of the sky (SMC, LMC, Galactic Bulge),
the variability survey used (MACHO, OGLE, MOA, EROS), use of infra-red
data (publicly available 2MASS or DENIS data, or own observations),
and the selection of the variables (unbiased, selection of Miras,
starting from spectroscopically known AGB stars).

The main results are very similar between these papers, and one of the
main results is shown in Figure~\ref{fig-PL-MC}. It shows different
sequences labelled ``ABCD''. Sequences $A$ and $B$ are further
split-up relative to the tip of the RGB. Sequence $B$ is now
often considered to be actually 2 sequences (e.g. Soszynski et al., 2004,
Fraser et al., 2005).

Selecting on amplitude selects stars in different boxes. Selecting the
largest amplitudes results in a nice Period-Luminosity relation in Box
$C$ that can be identified with the classical Mira variables. They are
almost certainly pulsating in the fundamental mode (Wood 1999, 2000,
Fraser et al. 2005). Selecting progressively smaller amplitude cuts
populates more and more Box $B$ and then $A$. These stars are thought
to be overtone pulsators (1st, 2nd, 3rd) although theoretical
$PL$-relation do not fit the observed sequences exactly (e.g Fraser et
al. 2005).

The big surprise to come out of these observations was sequence $D$.
Stars with a period in this box have multiple periods with the shorter
period in one of the other boxes. Examples of the light curves are
shown in Figure~\ref{Fig-D}. The origin of these, so-called, Long
Secondary Periods (LSPs) is still a mystery. Several explanations have
been proposed, $g^+$ non-radial modes, binary companions, rotating
prolate spheroids, episodic dust ejection, star spot cycles, but none
is satisfactory (Olivier \& Wood 2003, Wood et al. 2004).

\begin{figure}[!t]
\plottwo{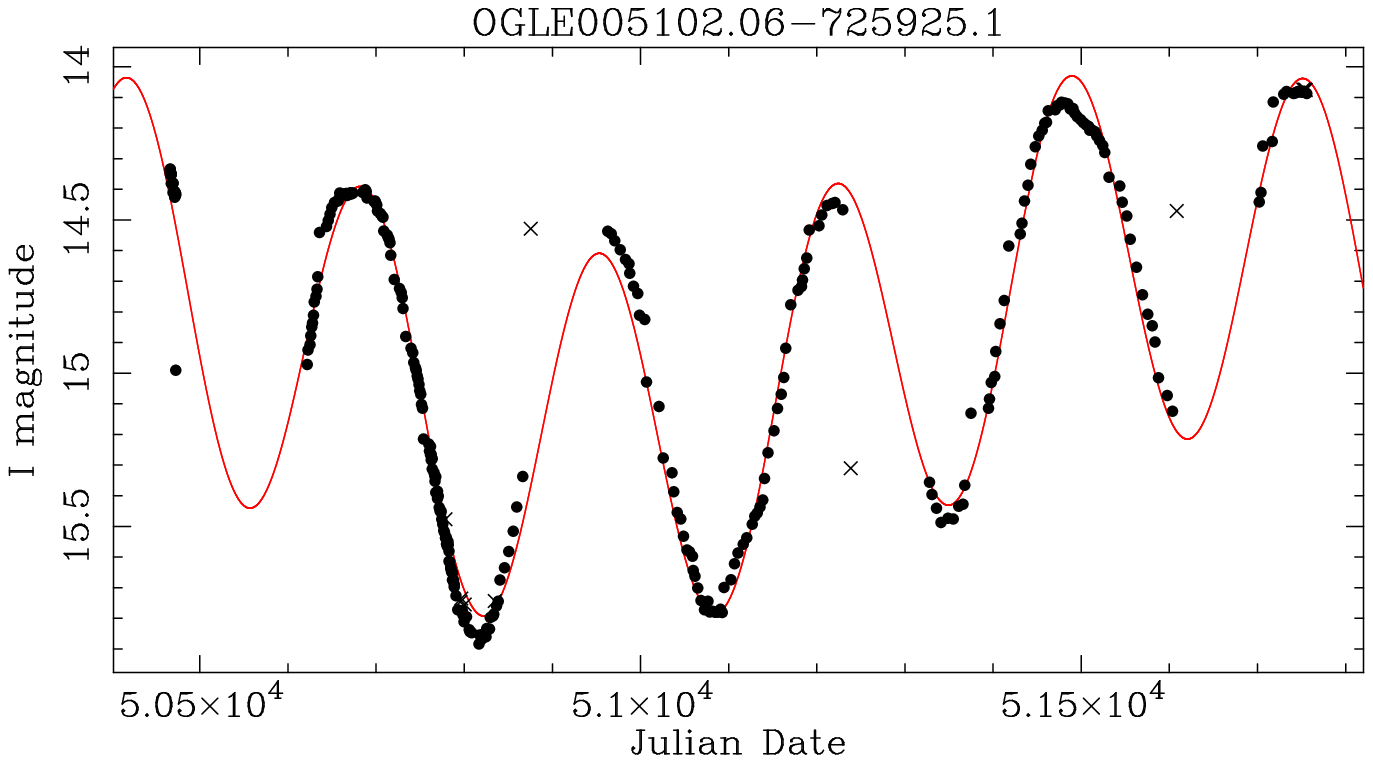}{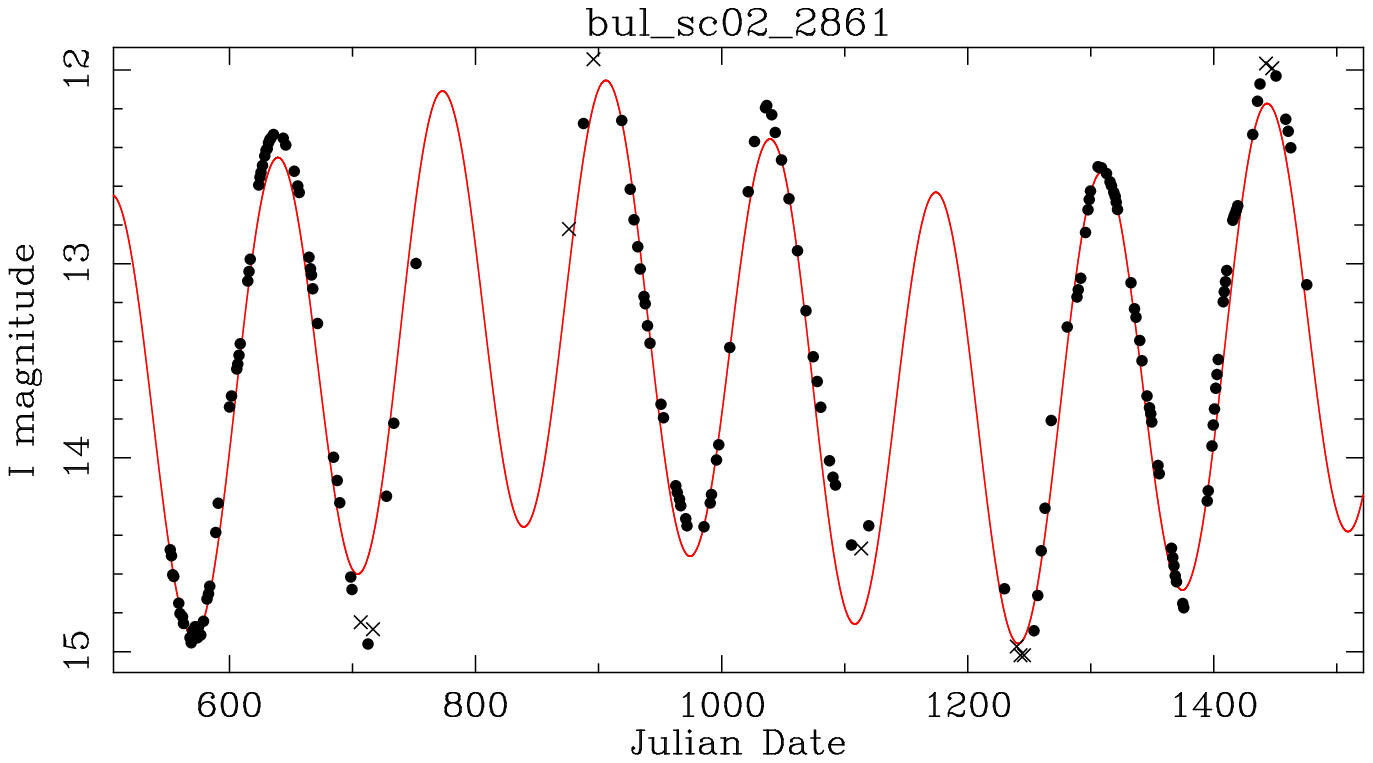}
\caption[]{
Example of LPVs with multiple periods one of which is in Box $D$. In
the left panel an LPV from the SMC with periods of 267 and 1300 days
(from Groenewegen 2004), in the right panel an example from the
Galactic Bulge with periods of 134 and 700 days (from Groenewegen \&
Blommaert, 2005).
}
\label{Fig-D}
\end{figure}

\begin{figure}[!t]
\plotone{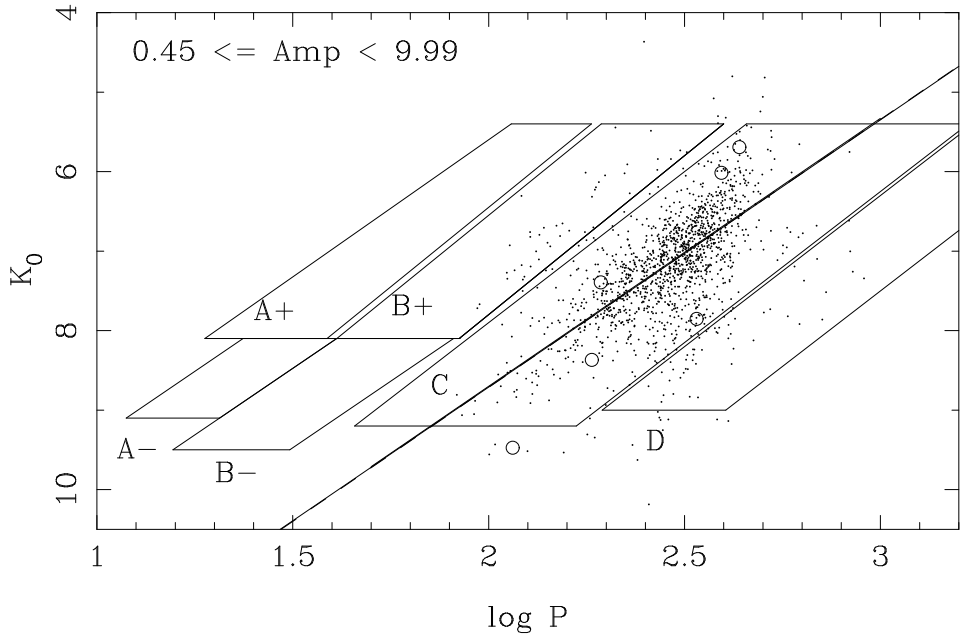}
\caption[]{
$K$-band $PL$-relation in the direction of the Galactic Bulge for periods with 
an $I$-band amplitude larger than 0.45 mag and $(J-K)_0 < 2.0$. 
Known M-stars are indicated by open circles.
The line indicates the $PL$-relation: $m_{\rm K}= -3.37  \log P + (15.47 \pm 0.03$).
From Groenewegen \& Blommaert (2005).
}
\label{Fig-PL-BUL}
\end{figure}

\subsection{The Galactic Bulge}

Table~1 includes the few works that have used the results of the
micro-lensing surveys to study variables in the direction of the
Galactic Bulge (GB). They extend previous classical works on Bulge
variable stars, like those of Lloyd Evans (1976), Glass \& Feast
(1982), Whitelock et al. (1991), Glass et al. (1995), Alard et
al. (1996) and Glass et al. (2001).

The work by Groenewegen \& Blommaert (2005, GB05) appears to be the
first paper to use the full OGLE-{\sc ii} database to study classical
Mira variables ($I$-band semi-amplitude larger than 0.45 magnitudes),
and some details of this study are presented here.

Figure~\ref{Fig-PL-BUL} presents the $K$-band $PL$-relation for the
2691 Mira variables identified in the OGLE database after correlation
with the 2MASS and DENIS. The mean $PL$-relation is also shown. 

\begin{figure}[!t]
\plotone{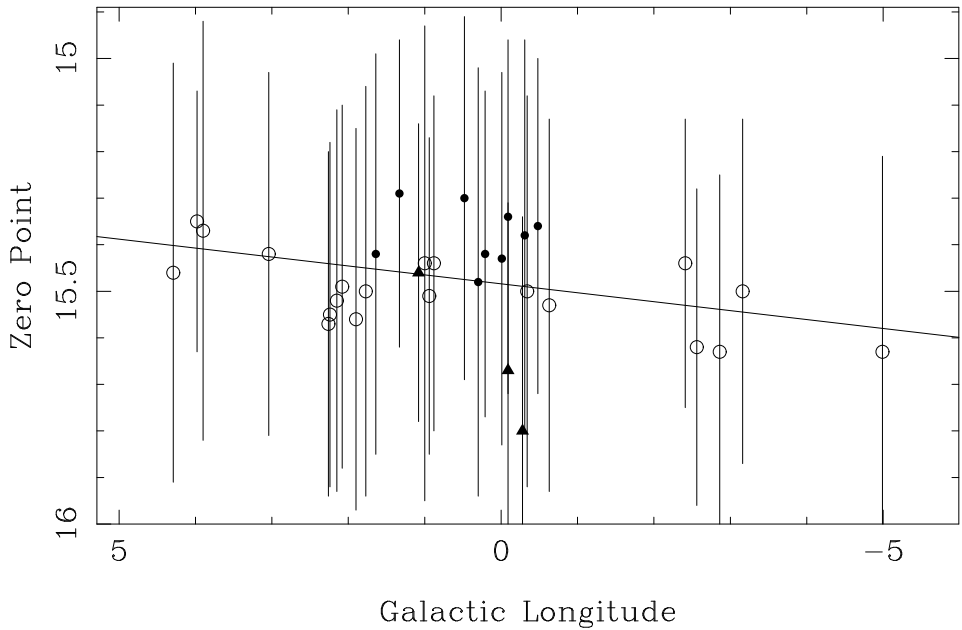}
\caption[]{
Zero point of the Mira $K$-band $PL$-relation in the Galactic Bulge as
a function of longitude for the inner fields with $\mid l \mid <5$
degrees. Galactic latitudes below $-4.0\deg$ are indicated by filled
triangles, those larger than $-2.6\deg$ by filled circles, and the
remaining by open circles. Error bars are also plotted. The line
represents a linear least-squares fit. }
\label{Fig-ZP}
\end{figure}

With this slope fixed, one can now determine the zero point (ZP) of
the $PL$-relation for each OGLE field separately. In Figure~\ref{Fig-ZP} 
the ZP is plotted against Galactic longitude for the inner OGLE
fields.  There is a significant slope indicating the presence of the
well-known bar. 

Simulations have been carried out using the basic disk \& bulge model
of Binney et al. (1997), with the main parameter being the angle
between the major axis of the bar and the line-of-sight towards the
Galactic Centre. It turns out that for angles of 43 and 79 degrees one
can account for the observed slope in Figure~\ref{Fig-ZP}, but only
the model with an angle of 43 degrees can also account for the
observed number of stars.

The preferred value of $\phi = 43\deg$ is in agreement with the
values of about $45\deg$ by Whitelock (1992), based on 104 IRAS
detected Mira variables, and the preferred value of $46$ degrees by
Sevenster et al. (1999), based on an analysis of OH/IR stars in the
inner Galaxy.

Other values in the literature are usually much larger, between 60 and
80 degrees: Dwek et al. (1995) and Binney et al. (1997), based on
COBE-DIRBE data, Stanek et al. (1997), based on bulge red clump stars,
Robin et al. (2003) and Picaud \& Robin, based on colour-magnitude
fitting.  Sevenster et al. (1999), however, argues that these values
are commonly found when no velocity data is available, the longitude
range is too narrow or when low latitudes are excluded.  It is also
possible that these studies are tracing other populations, which may
be differently distributed than the Miras. Whitelock et al.  and
Sevenster et al. do use populations closely related to the Mira stars
and find an angle of the bar close to the one we derive. \\

\begin{figure}[!t]
\begin{minipage}{0.47\textwidth}
\resizebox{\hsize}{!}{\includegraphics{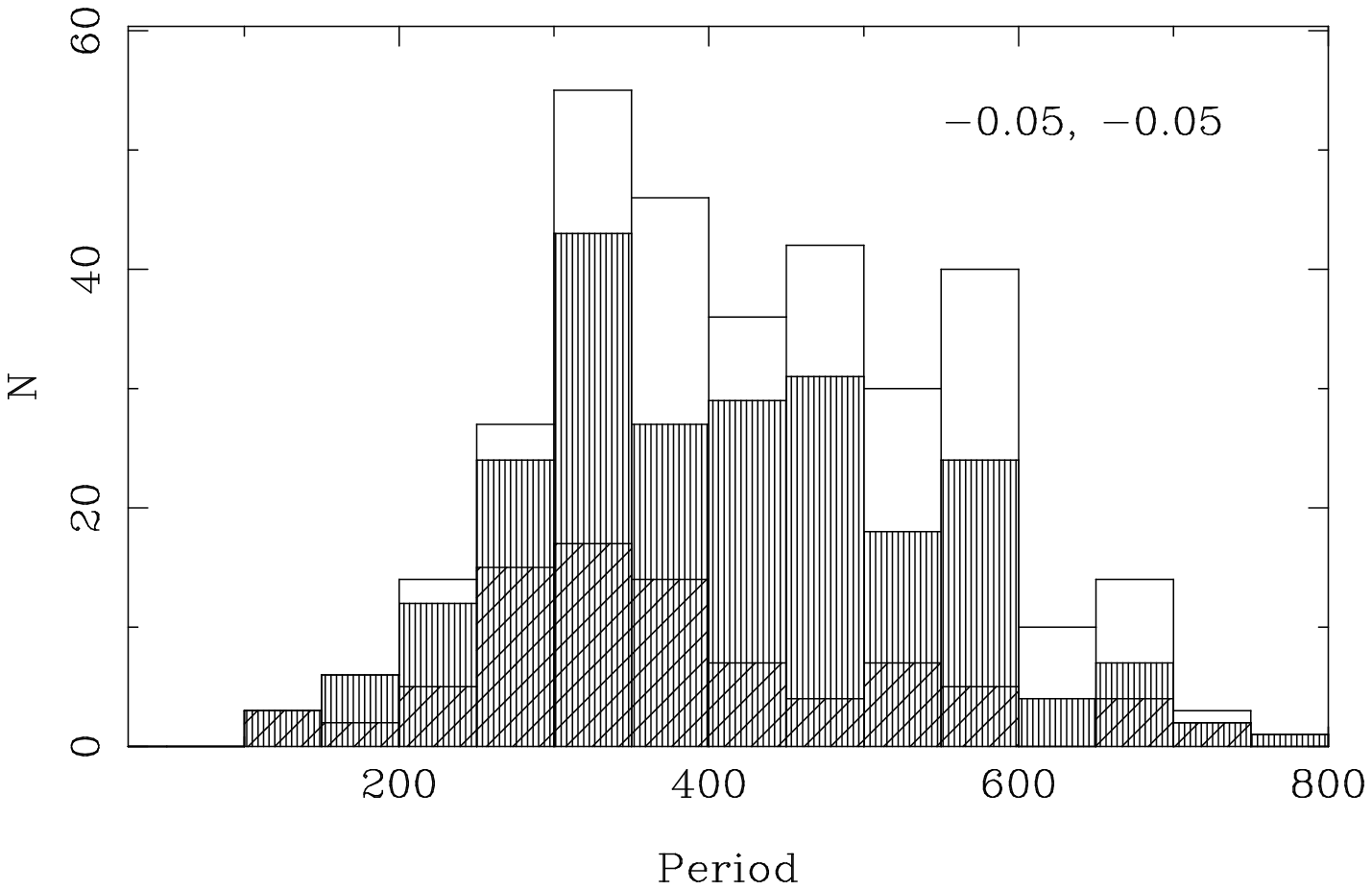}}
\end{minipage}
\hfill
\begin{minipage}{0.47\textwidth}
\resizebox{\hsize}{!}{\includegraphics{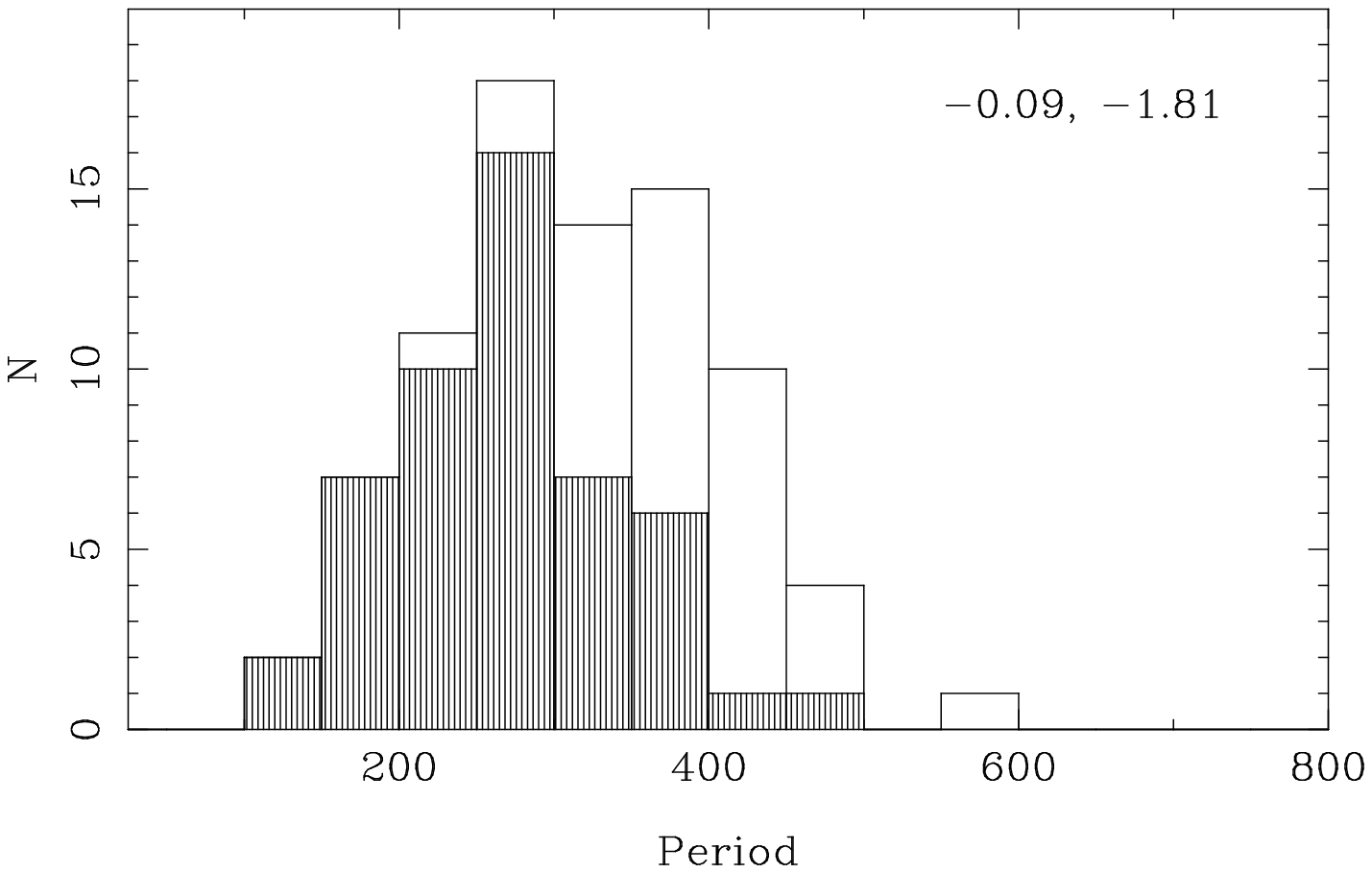}}
\end{minipage}

\begin{minipage}{0.47\textwidth}
\resizebox{\hsize}{!}{\includegraphics{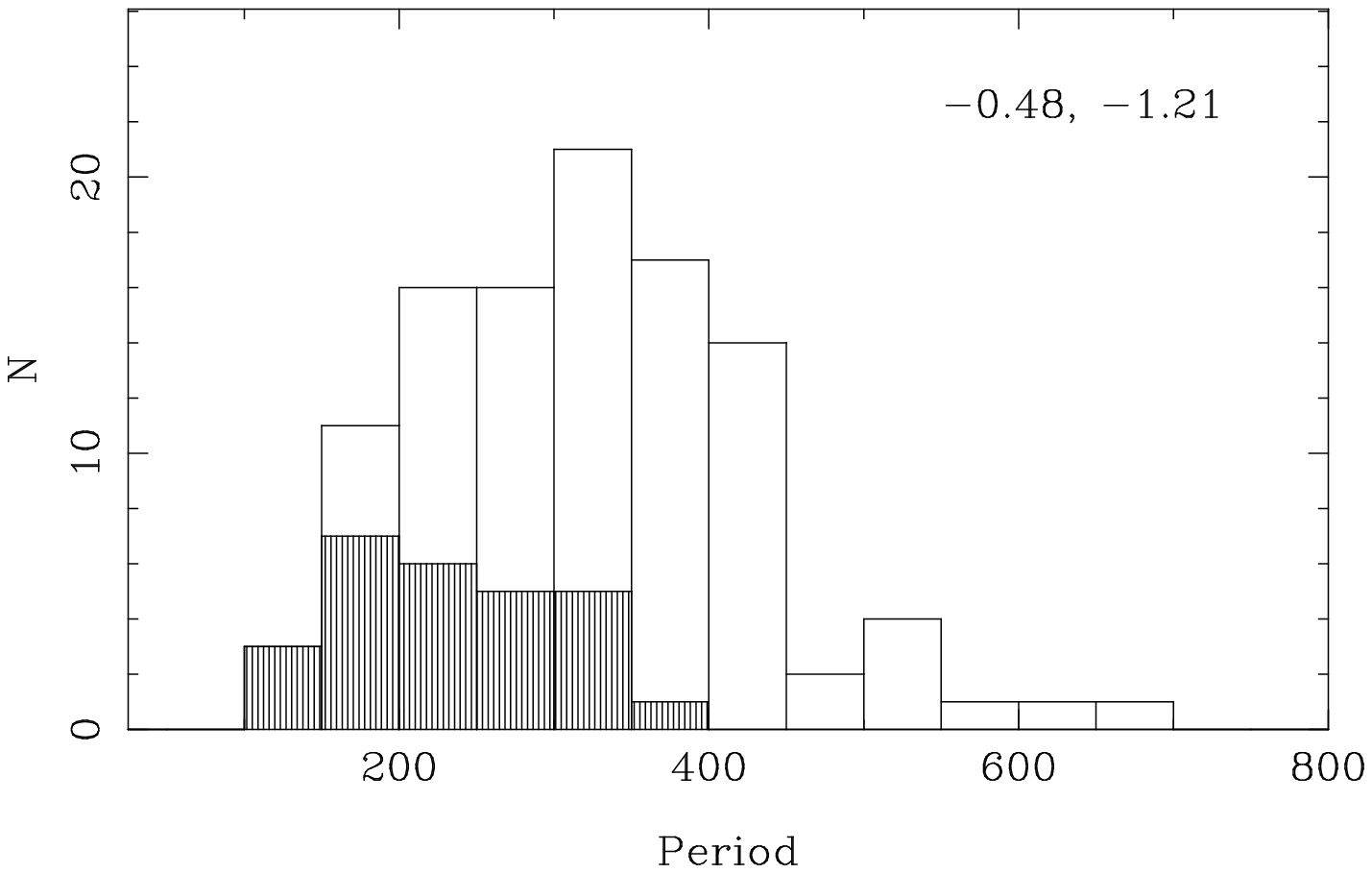}}
\end{minipage}
\hfill
\begin{minipage}{0.47\textwidth}
\resizebox{\hsize}{!}{\includegraphics{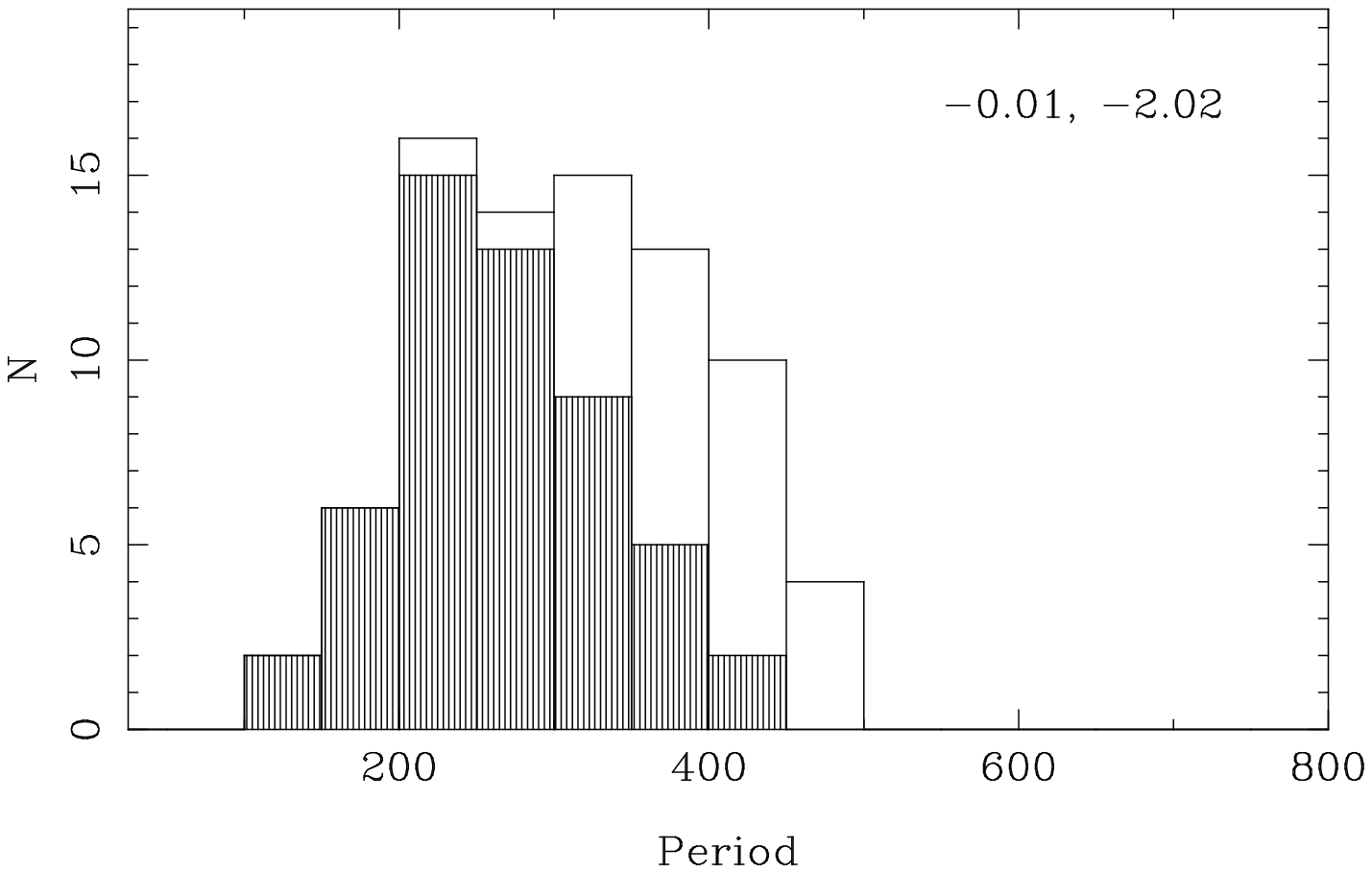}}
\end{minipage}

\begin{minipage}{0.47\textwidth}
\resizebox{\hsize}{!}{\includegraphics{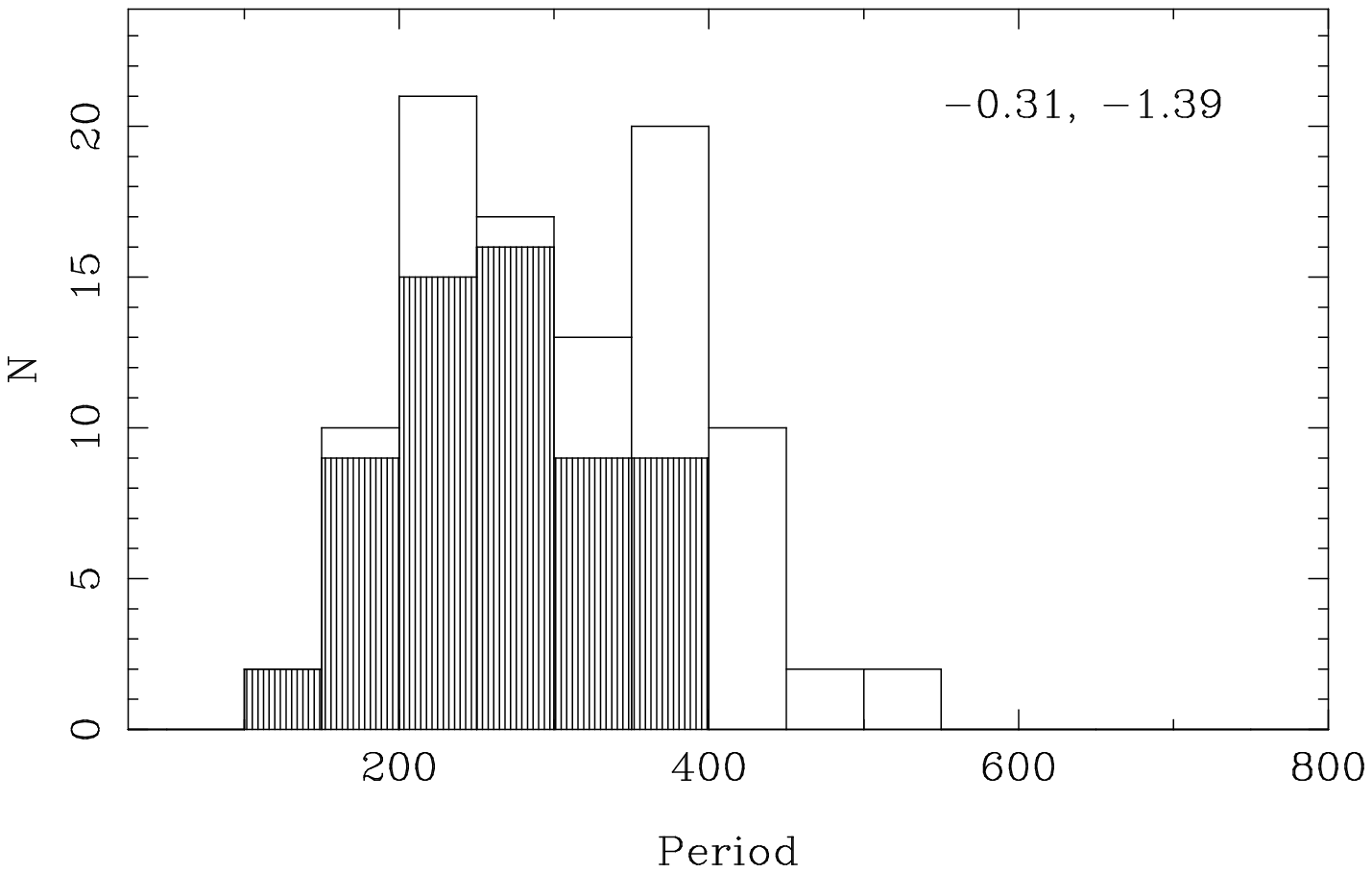}}
\end{minipage}
\hfill
\begin{minipage}{0.47\textwidth}
\resizebox{\hsize}{!}{\includegraphics{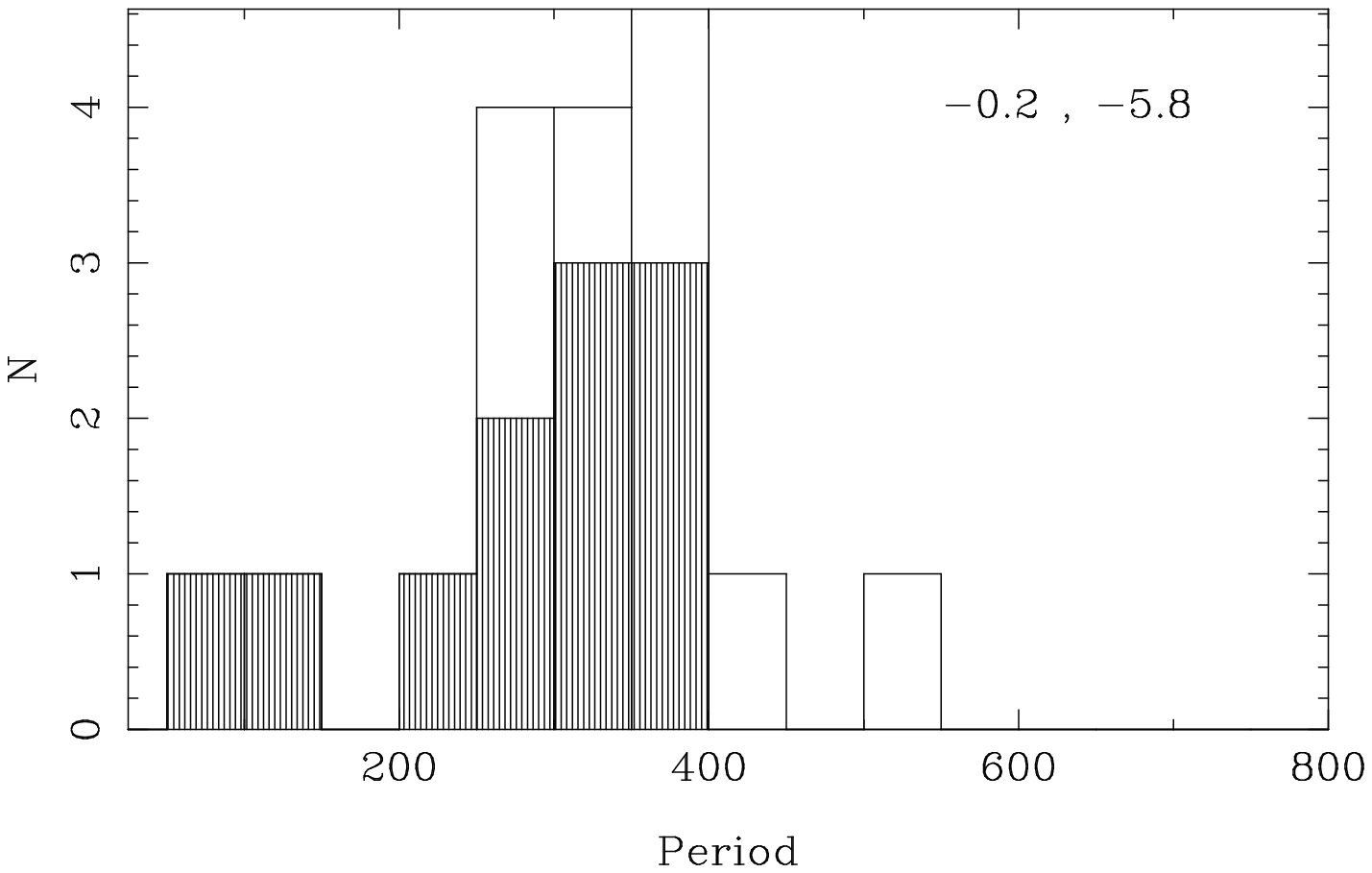}}
\end{minipage}

\caption[]{
Galactic Bulge Mira period distribution for 6 fields with similar
longitudes but a range in latitudes (as indicated in the top right
corner). For the field at $b \sim -5.8$ degree, OGLE fields 6 and 7
have been combined.  For the shaded histograms only stars with
$(J-K)_0 < 2.0$ have been included.  The field at $(-0.05, -0.05)$ is
based on Glass et al. (2001), see main text for details. The histogram
with slanted hatching is for the reddening by Schultheis et al. (1999) 
for stars in this field, the shaded histogram for the adopted
reddening which is 1.35 times larger. From Groenewegen \& Blommaert (2005).
}
\label{Fig-PerField}
\end{figure}

\noindent
Another way to look at this dataset is shown in Figure~\ref{Fig-PerField} 
where the period distribution is shown for OGLE-fields at similar
latitudes but different latitudes. To add a field even closer to the
GC than surveyed by OGLE the data in Glass et al. (2001, 2002) is
considered on a field centered on $l = -0.05\deg, b = -0.05\deg$. They present
the results of a $K$-band survey of 24 $\times$ 24 arcmin$^2$ for LPVs
down to $K \sim 12.0$. 

A Kolmogorov-Smirnov test indicates that the fields at $-1.2\deg$ and
lower are statistically identical, and very different from the inner
field, where a tail of longer periods is observed.

\begin{figure}[!t]

\begin{minipage}{0.48\textwidth}
\resizebox{\hsize}{!}{\includegraphics{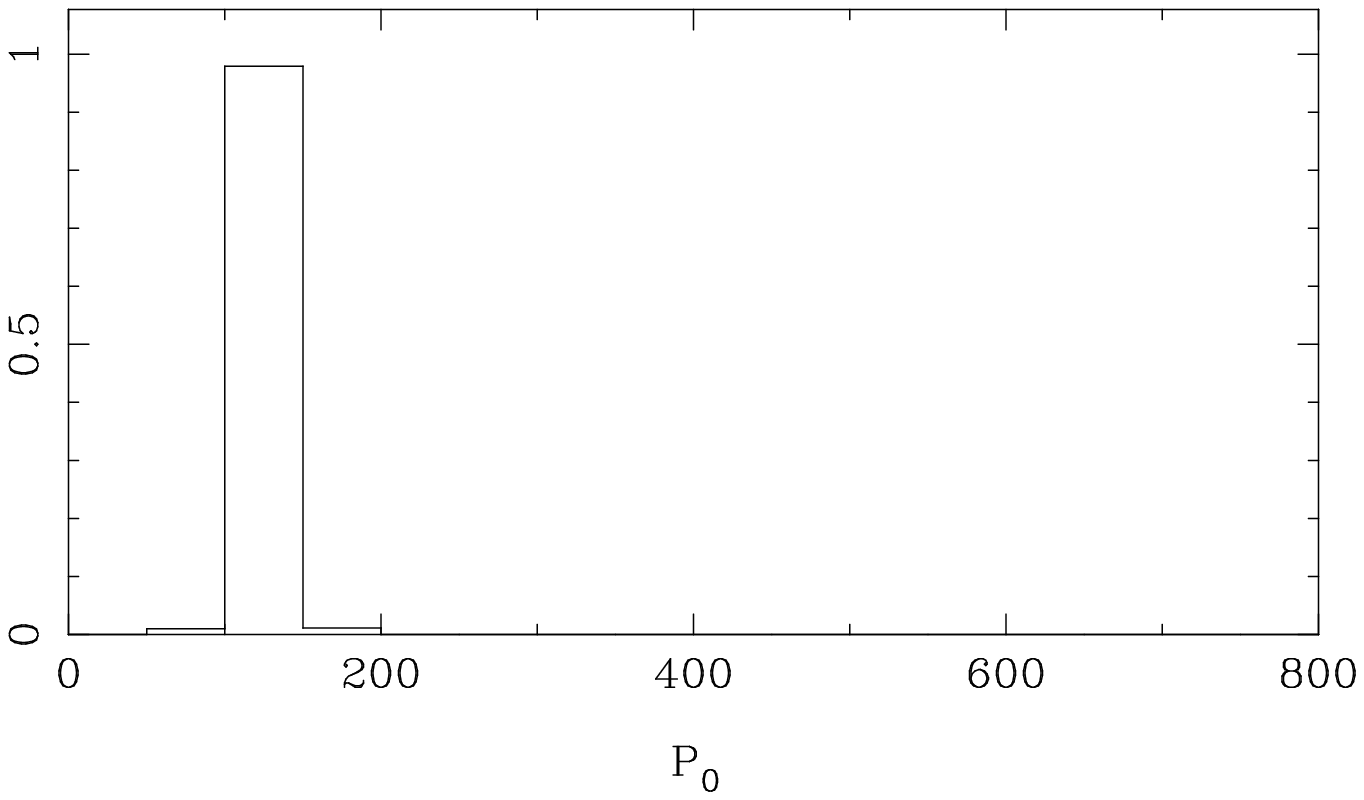}}
\end{minipage}
\hfill
\begin{minipage}{0.48\textwidth}
\resizebox{\hsize}{!}{\includegraphics{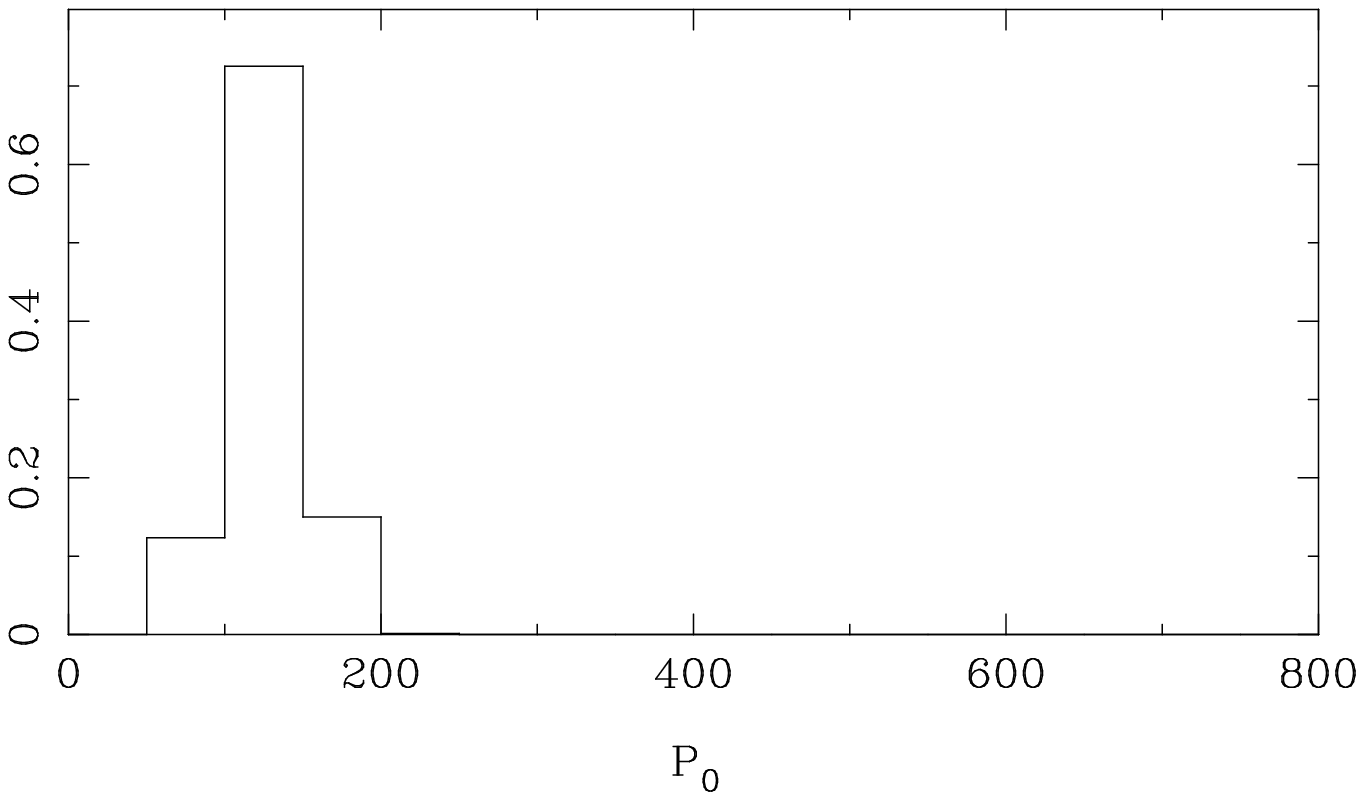}}
\end{minipage}

\begin{minipage}{0.48\textwidth}
\resizebox{\hsize}{!}{\includegraphics{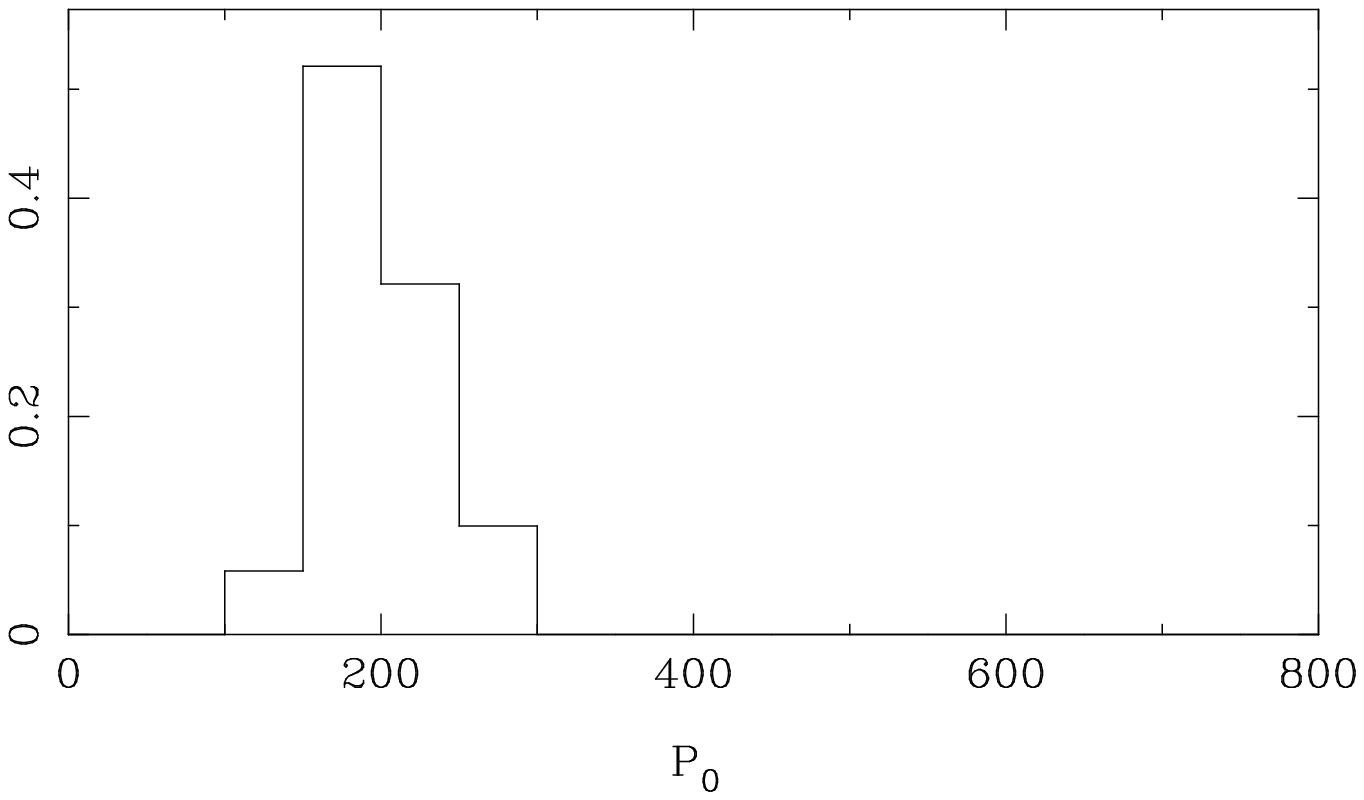}}
\end{minipage}
\hfill
\begin{minipage}{0.48\textwidth}
\resizebox{\hsize}{!}{\includegraphics{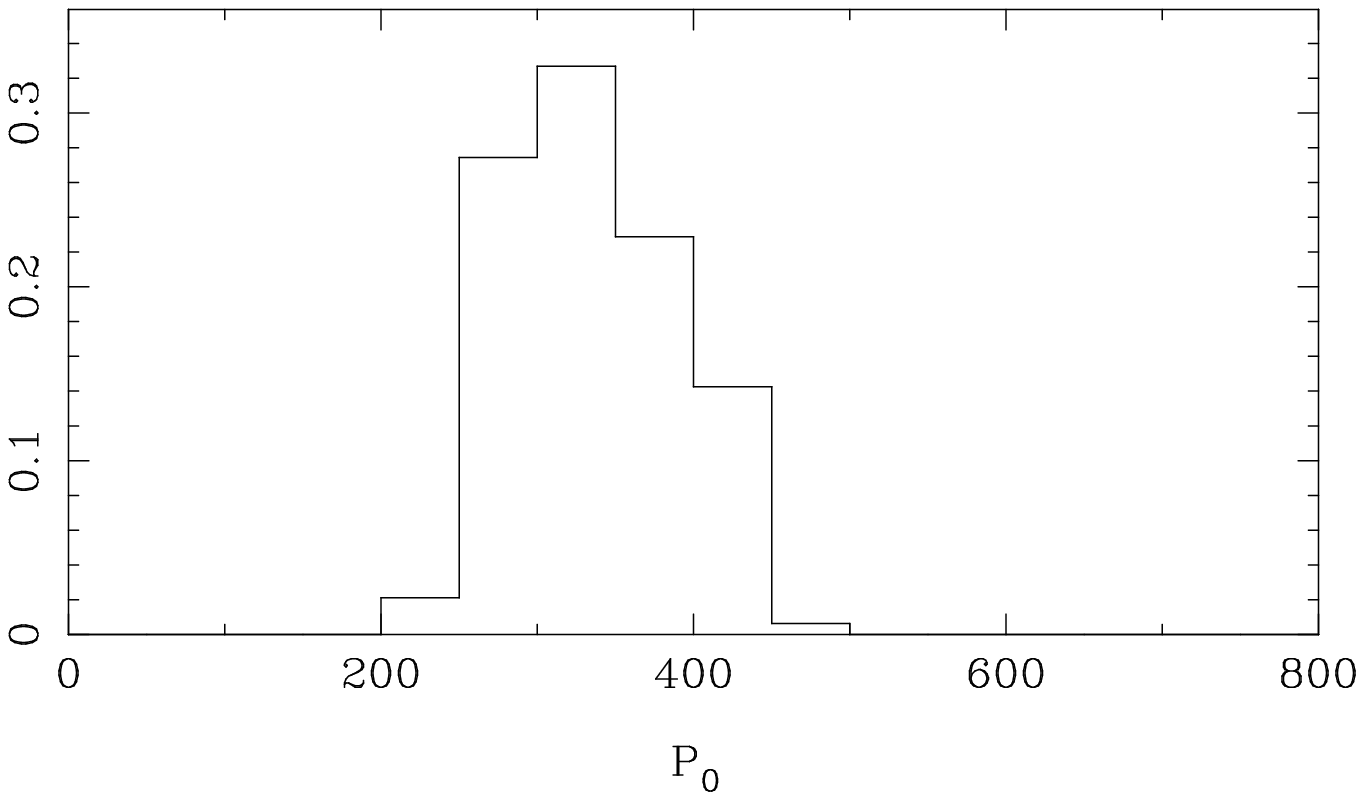}}
\end{minipage}

\begin{minipage}{0.48\textwidth}
\resizebox{\hsize}{!}{\includegraphics{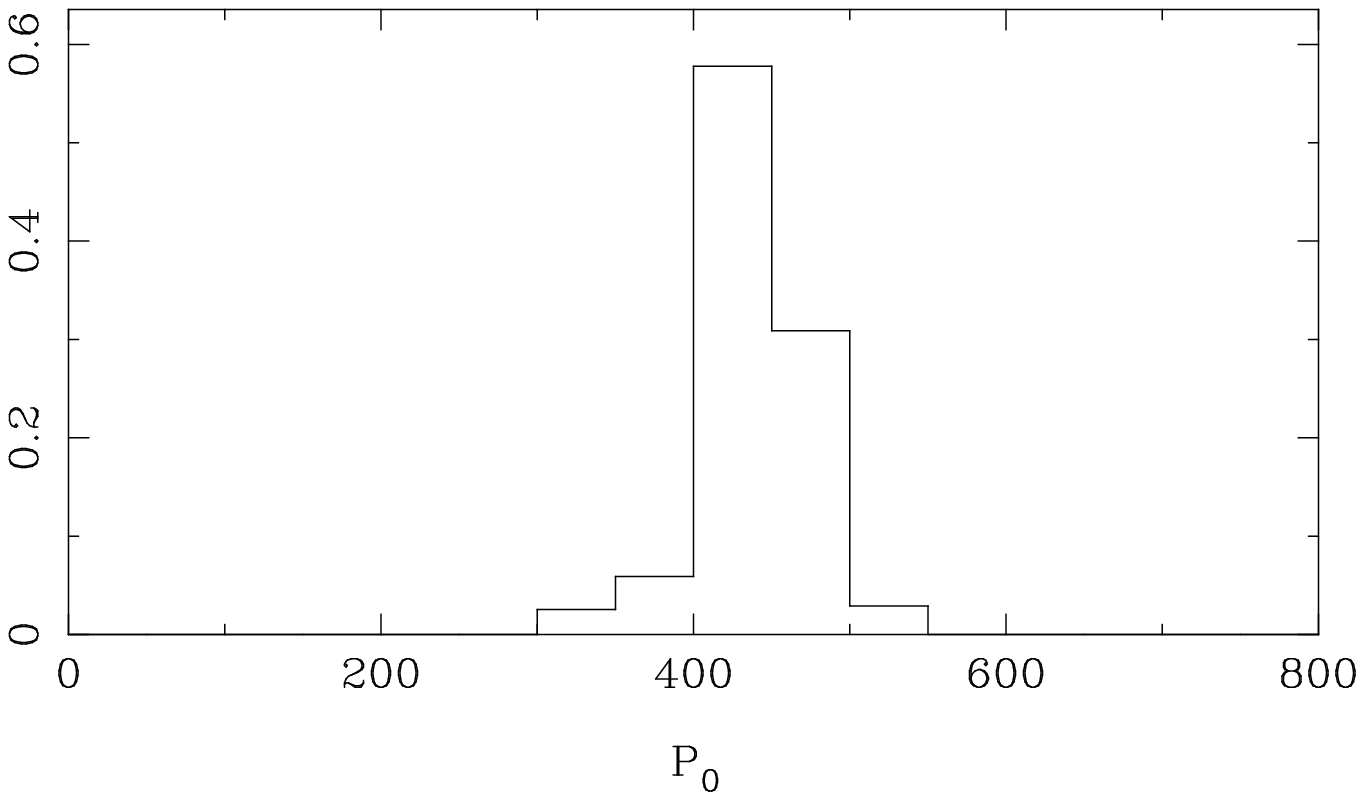}}
\end{minipage}
\hfill
\begin{minipage}{0.48\textwidth}
\resizebox{\hsize}{!}{\includegraphics{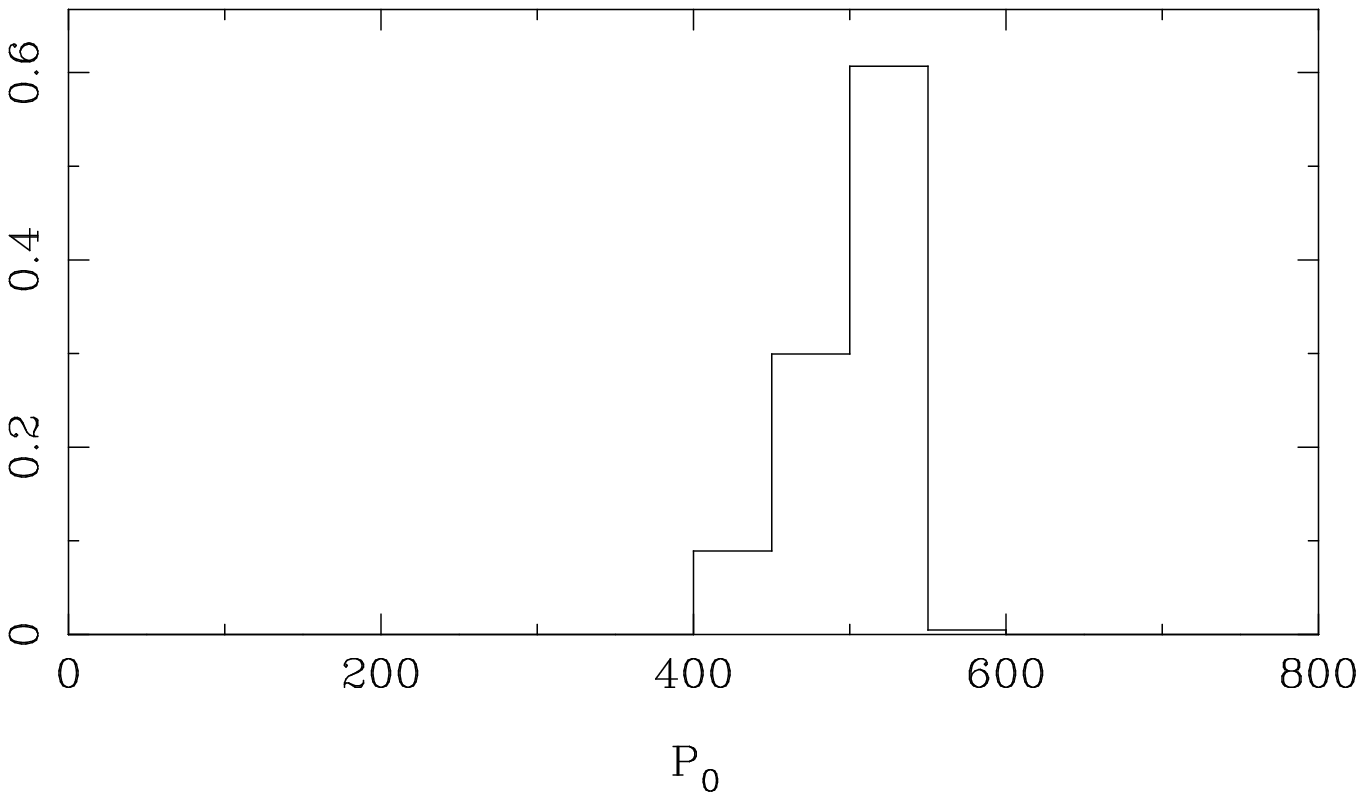}}
\end{minipage}

\caption[]{
Theoretical period distribution of optically visible stars inside the observed
instability strip for masses 1.1, 1.2, 1.5, 2.0, 2.5, 3.0 M$_\odot$ (left to
right, top to bottom).
}
\label{Fig-PerDistr}
\end{figure}

To quantify the nature of the Mira Bulge population, synthetic AGB
evolutionary models have been calculated. In brief, the synthetic AGB
code of Wagenhuber \& Groenewegen (1998) was finetuned to reproduce
the models of Vassiliadis \& Wood (1993) for $Z$ = 0.016 and then
extended to more initial masses and including mass loss on the
RGB. For several initial masses the fundamental mode period
distribution was calculated for stars inside the observed instability
strip and when the mass loss was below a critical value to simulate
the fact that they should be optically visible. The results are shown
in Figure~\ref{Fig-PerDistr}.

From the comparison of the observed period distribution for fields
more than $1.2\deg$ away from the galactic centre with the
theoretical ones, we deduce that the periods can be explained with a
population of stars with Main Sequence masses in the range of 1.5 to
2.0 M$_\odot$. A possible extension to smaller masses is possible, but not
necessary to explain the periods below 200 days.  To explain the
excess periods in the range of 350-600 days observed closer to the
centre we need initial masses in the range 2.5 - 3 M$_\odot$.  The
presence of more massive stars in the inner field at $b$ = $-0.05\deg$ 
cannot be excluded, as it turns out that for more massive stars the
optically visible Mira phase is essentially absent.

The formation history of the Bulge is still a matter of debate. In
several works like in Kuijken \& Rich (2002) and recently in Zoccali
et al. (2003), the bulge is considered to be old ($>10$~Gyr) and
formed on a relatively short timescale ($< 1$~Gyr) (e.g. Ferreras et
al. 2003).  The bulge Miras do not fit in this picture as, according
to the analysis in GB05, they are considerably younger.

Their results agree more with the analysis of the infrared ISOGAL survey
discussed in van Loon et al. (2003). They conclude that the bulk of
the bulge population is old (more than 7~Gyr) but that a fraction of
the stars is of intermediate age (1 to several Gyr).  The Miras in GB05
study can thus be considered as the intermediate age population seen
in their analysis. 

If indeed the bulk of the bulge population is old and formed quickly
and if the Miras are of intermediate age, then the Miras must be
representatives of a population which was added at a later stage and
it is unclear how it relates to the overall bulge.  An interesting
scenario suggested in Kormendy \& Kennicutt (2004) is the one in which
a secondary bulge or also called pseudo-bulge forms within an old
bulge. Such a process would be connected to the presence of a ``bar''
which would add ``disky'' material into the old classical bulge.  The
Miras are indeed situated in a bar-structure as was discussed earlier.

\section{AGB stars in the Local Group using narrow-band filter surveys}

This technique uses the specific characteristic of late-type M-stars,
where strong TiO bands develop, and C-stars, with C$_2$ and CN
molecular bands.  First introduced by Palmer \& Wing (1982) and then
applied by Richer et al. (1984) and Aaronson et al. (1984) the method
typically uses two broad-band filters from the set $V,R,I$, and two
narrow-band filters near 7800 and 8100 \AA, which are centred on a
CN-band in carbon stars (and near-continuum in oxygen-rich stars), and
a TiO band in oxygen-rich stars (and continuum in C-stars),
respectively. In an [78-81] versus $[V-I]$ (or $[R-I]$) colour-colour
plot, carbon stars and late-type oxygen-rich stars clearly separate
redwards of $(V-I) \approx$ 1.6.  For an illustration of this, see
Cook \& Aaronson (1989) or Nowotny \& Kerschbaum (2002).

Originally, the applications were limited mainly because of the small
format CCDs and the use of relatively small telescopes. However the
last about 5 years have seen a remarkable revival through the use of
these narrow-band filters using wide-field imagers on bigger
telescopes (2-4m class).

At present a large fraction of LG galaxies have been
surveyed, at least partially, using these narrow-band filters. For
recent reviews see Azzopardi (1999) and Groenewegen (1999, 2002, 2005). 
The most recent works not listed in these reviews are the surveys by:

\begin{itemize}

\item Kerschbaum et al. (2004) who surveyed a 6.5\arcmin\ $\times$
6.5\arcmin\ field covering almost all of And {\sc ii} and identifying
7 C-stars.

\item Demers et al. (2004) who identified 676 carbon stars in a 42\arcmin\
$\times$ 28\arcmin\ field centered on IC10.

\item Battinelli \& Demers (2005a) who surveyed a 450 arcmin$^2$ field
about 40 kpc along the southern major axis of M31 finding only a
handfull of C-stars, concluding they reached the edge of the M31 disk,
at least as defined by an intermediate age population.

\item Harbeck et al. (2005) who found one candidate carbon star in
And {\sc ix}.

\item Rowe et al. (2005) who surveyed a 74\arcmin\ $\times$ 56\arcmin\
field centered on M33 finding 7936 C-stars.

\end{itemize}

\begin{figure}[!t]

\plotone{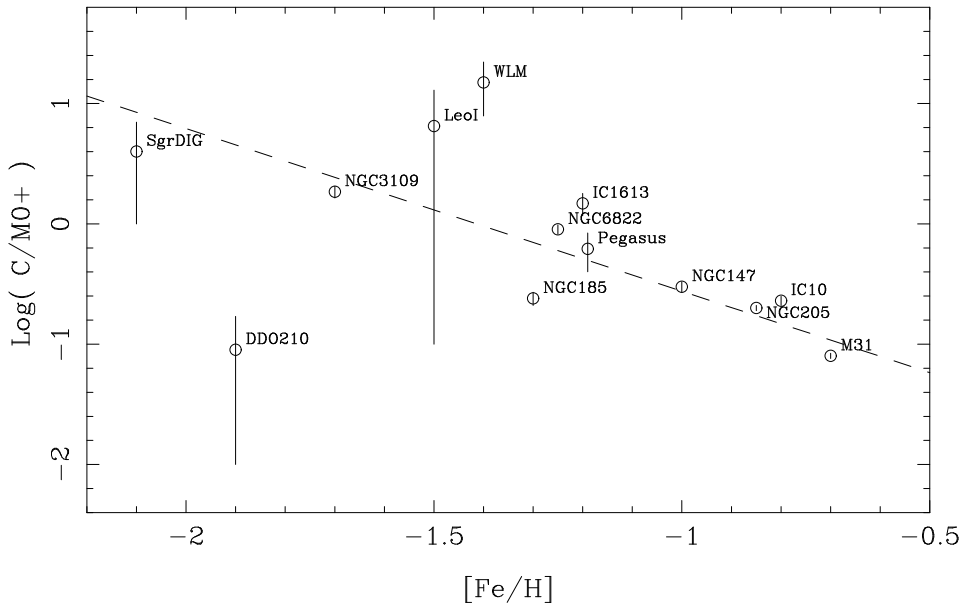}

\caption[]{Log (number Carbon stars / number  M0+ -stars) versus metallicity. 
Data taken from Battinelli \& Demers (2005b). The line is
a weighted least-squares fit to the data: 

$\log$ (C/M0+) = $(-1.35 \pm 0.20)$ [Fe/H] + $(-1.92 \pm 0.20)$.}

\label{MON5}
\end{figure}

A quantitative theoretical interpretation or understanding of this
data is clearly lacking. At the lowest level of interpretation these
data indicate a relation between the ratio of C-to-M stars and (mean)
metallicity in the Galaxy, as shown in Figure~\ref{MON5} (data from
Battinelli \& Demers, 2005b).

\begin{figure}[!t]

\plotone{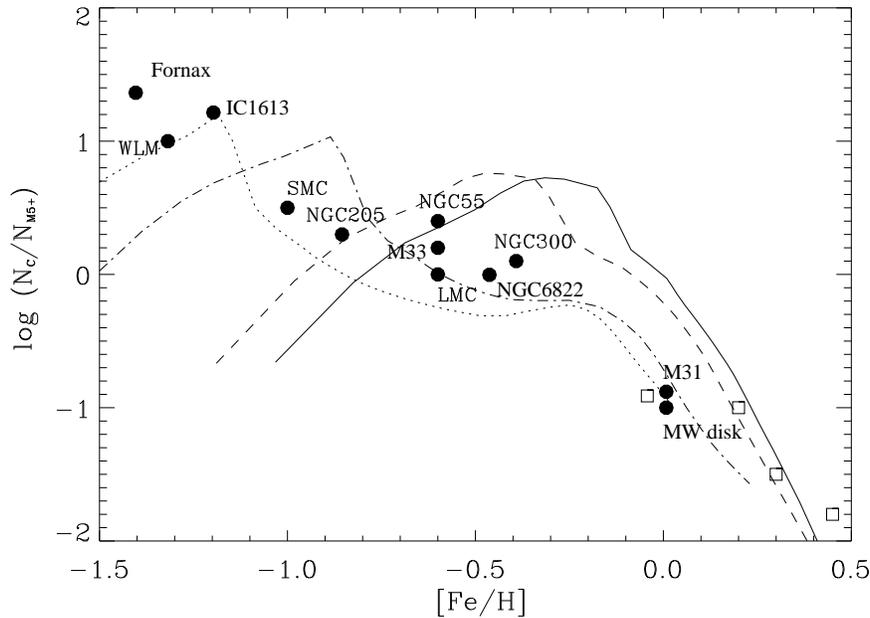}

\caption[]{Figure 8 from Mouhcine \& Lan\c{c}on (2003) showing Log
(number Carbon stars / number M5+ -stars) versus metallicity using
data points from Groenewegen (1999). Lines indicate model predictions
assuming typical SFRs, characteristic of Sa (solid), Sb (sahed), Sc
(dot-dash) and Irr (dotted) type galaxies. 
\vspace{-3mm}
}
\label{Must}
\end{figure}

Mouhcine \& Lan\c{c}on (2003, see Figure~\ref{Must}) present
evolutionary population synthesis models, including chemical
evolution, with special focus on intermediate age populations. Their
models are the first that are able to account qualitatively for the
observed trend in Figure~\ref{MON5} adopting `typical' Star Formation
Histories (SFHs) for Sa, Sb, Sc and Irr Hubble type galaxies.  The AGB
phase is included through a semi-analytical treatment of the third
dredge-up, with efficiency parameters set to values that have been
determined in other studies to fit the LMC carbon star LF and C/M
ratio.

With SFH now (becoming) available for most LG galaxies (see Dolphin,
this proceedings) one could investigate for individual galaxies
whether these SFHs predict the observed number of M- and C-type AGB
stars.  Repeating this for a large set of galaxies covering a range of
metallicities would possible better constrain individual SFHs for
intermediate ages, as well lead to a comprehensive picture of dredge-up 
and mass loss efficiency as a function of metallicity in AGB stars.






\end{document}